\documentclass[usenatbib]{mnras}

\usepackage{newtxtext}
\usepackage[varvw]{newtxmath} 
\usepackage{graphicx}          
\usepackage{microtype}
\usepackage{cleveref}

\graphicspath{{./figures/}}
\hypersetup{linkcolor=red}     % red equations

\newcommand{\sn}{\mathcal{S/N}}

\def\equationautorefname~#1\null{equation~(#1)\null}

\title[JWST Ultramassive Galaxy Sample -- II]{The JWST Ultramassive Galaxy Sample -- II. Supermassive Black Hole Masses for 8 Extreme Early-Type Galaxies via Jeans Anisotropic Modelling of High-Resolution Integral-Field NIRSpec Stellar Kinematics}
    
\author[M. Cappellari, D. D. Nguyen, and S. A. Kassin]{
Michele Cappellari$^{1}$\thanks{E-mail: michele.cappellari@physics.ox.ac.uk},
Dieu D. Nguyen$^{2}$\thanks{E-mail: dieun@umich.edu} and Susan A. Kassin$^{3}$
\\
$^{1}$Sub-Department of Astrophysics, Department of Physics, University of Oxford, Denys Wilkinson Building, Keble Road, Oxford, OX1 3RH, UK\\
$^{2}$Department of Astronomy, University of Michigan, 1085 South University Avenue, Ann Arbor, MI 48109, USA\\
$^{3}$Space Telescope Science Institute, 3700 San Martin Drive, Baltimore, MD 21218, USA
}

\date{Submitted to MNRAS on 2026 September 18}

\pagerange{\pageref{firstpage}--\pageref{lastpage}} \pubyear{2026}

\begin{document}

\label{firstpage}
\maketitle

\begin{abstract}
We present central supermassive black hole (BH) mass measurements for the JWST Ultramassive Galaxy Sample, comprising 8 galaxies selected from a complete sample of the most massive early-type galaxies in the nearby Universe ($M_\star \gtrsim 2\times10^{12}\,\mathrm{M}_\odot$). Building upon the high-resolution Multi-Gaussian Expansion (MGE) surface brightness profiles and exquisite JWST/NIRSpec spatially resolved stellar kinematics introduced in Paper I, we employ the spectral Jeans Anisotropic Modelling (JAM) framework to reconstruct the nuclear stellar dynamics. Given the unique structural properties of these dry-merger-dominated systems, our Bayesian MCMC analysis explicitly tests both oblate and prolate intrinsic geometries. By formally marginalizing over classical dynamical degeneracies---including orbital velocity anisotropy, stellar mass-to-light ratios, and viewing inclinations---we extract robust, statistically rigorous BH masses. We demonstrate that the exceptional spatial resolution of JWST allows the JAM method to tightly constrain the central gravitational potential using nuclear kinematics alone. Crucially, the modelling reveals that not all ultramassive galaxies host ultramassive black holes: while the majority display the classic central velocity dispersion peak, notable exceptions emerge, including a very clear central dispersion drop in one galaxy and a subdued peak in another. These qualitative differences translate directly into distinct BH masses, indicating that BH growth at the extreme mass scale may be more stochastic than traditionally assumed, potentially influenced by three-body ejections during hierarchical mergers or gravitational recoil kicks from coalescing binaries. In the latter case, while the recoil velocity is unlikely to exceed the escape velocity of an ultramassive galaxy, the kicked BH may currently be displaced and wandering outside the compact NIRSpec field of view. The BH masses derived in this work successfully populate the historically sparse upper envelope of the galaxy mass hierarchy. These measurements serve as the critical dynamical foundation for Paper~III, where they are used to investigate extreme-mass BH scaling relations and test theoretical predictions regarding the impact of dissipationless mergers on galaxy-BH co-evolution.
\end{abstract}


\section{Introduction}
\label{sec:intro}

The co-evolution of supermassive black holes (BHs) and their host galaxies is a cornerstone of modern extragalactic astrophysics \citep{Kormendy2013review}. This interconnected growth is strongly supported by the tight empirical scaling relations observed between the BH mass, $M_{\rm BH}$, and the large-scale properties of the host spheroid, such as its stellar velocity dispersion, $\sigma_\star$ \citep[e.g.][]{Ferrarese2000, Gebhardt2000bh}, and total stellar mass, $M_\star$ \citep{Marconi2003, Haring2004}. While these relations are reasonably well-established for intermediate-mass systems, the extreme high-mass end remains poorly sampled.

\subsection{Black hole scaling relations at the top of the mass hierarchy}
\label{sec:intro_ultramassive}

In the first paper of this series \citep[hereafter Paper I]{Cappellari2026p1}, we introduced the JWST Ultramassive Galaxy Sample, 8 extreme early-type galaxies (ETGs) with $M_\star \gtrsim 2\times 10^{12}\,\mathrm{M}_\odot$ extracted from a full census of nearby ultramassive galaxies \citep{Nguyen2023}. These ultramassive galaxies represent a very special subset of objects: they reside at the very bottom of the hierarchical galaxy formation process. Because their late-time assembly is overwhelmingly dominated by dry, dissipationless mergers \citep{Oser2010, Naab2017}, studying them allows us to push our understanding of the entire hierarchical growth process to its limits. Crucially, dry mergers can substantially increase both $M_\star$ and $M_{\rm BH}$ while leaving $\sigma_\star$ comparatively unchanged, potentially causing a transition where $M_{\rm BH}$ scales more directly with $M_\star$ rather than $\sigma_\star$ \citep{Krajnovic2018channels}.

To test this hypothesis, we must accurately measure the BH masses in these extreme systems. There are various methods available to measure $M_{\rm BH}$, including the dynamics of megamasers \citep[e.g.,][]{Miyoshi1995} and the kinematics of cold or ionized gas \citep[e.g.,][]{Barth2001, Davis2013}. However, for ultramassive galaxies, the only general technique is to use stellar dynamics. These mature systems are typically devoid of regular, rotating discs of ionized or cold molecular gas that are required for reliable gas-dynamical modelling, although there are notable exceptions, such as the similarly massive galaxy M87 \citep[e.g.,][]{Harms1994, Macchetto1997}.

From a stellar-dynamical perspective, ultramassive galaxies possess unique features that must dictate our modelling approach. Above a critical mass threshold of $M_{\rm crit} \approx 2\times 10^{11}\,\mathrm{M}_\odot$, galaxies transition dynamically: they tend to be ``slow rotators'' \citep{Emsellem2011p3, Cappellari2013p20} and are generally weakly triaxial \citep{Weijmans2014p24, Li2018shapes} \citep[see reviews by][]{Cappellari2016, Cappellari2026}. However, when climbing further up the mass ladder to $\gtrsim 10^{12}\,\mathrm{M}_\odot$---which is precisely the regime explored by our sample---there is compelling evidence that this weak triaxiality transitions toward strongly prolate shapes \citep{Krajnovic2018prolate}. This fascinating structural shift was recently beautifully illustrated by \citet{Krajnovic2026}. The presence of these complex, prolate-like geometries has important consequences for our dynamical modelling choices.

\subsection{Complementary approaches to stellar-dynamical modelling}
\label{sec:intro_modelling_methods}

Stellar-dynamical BH masses are commonly measured using two complementary classes of dynamical models: Schwarzschild orbit-superposition models \citep{Schwarzschild1979} and Jeans models \citep{Jeans1915}. The two approaches occupy different points in the trade-off between model flexibility and statistical regularization. 

Fully triaxial Schwarzschild models \citep[e.g.,][]{vandenBosch2008, Vasiliev2020, LingZhu2020, Neureiter2021, Quenneville2021, Thater2022dynamite} provide the least parametrically restrictive description of the stellar orbital distribution and are therefore particularly valuable when sufficiently spatially-extended and high $S/N$  kinematic data are available to constrain that freedom. The JAM framework \citep{Cappellari2008, Cappellari2020, Cappellari2026jam} deliberately occupies a different point in this trade-off: it replaces a non-parametric orbit distribution with a parametrized description of the velocity anisotropy, thereby regularizing the inverse problem and allowing the gravitational potential to be constrained from substantially less extensive kinematic data.

Jeans models formally possess a mass--anisotropy degeneracy \citep{Binney1982, Gerhard1993}: different combinations of gravitational potential and orbital anisotropy can reproduce similar projected second moments. In practical applications, however, the anisotropy is not unconstrained. Decades of orbit-based dynamical studies have established strong empirical regularities in the orbital structure of real galaxies \citep[see review and fig.~10 in][]{Cappellari2026}. JAM exploits this information explicitly through physically motivated anisotropy parametrizations and priors. Conversely, the much greater orbital freedom of Schwarzschild models must be constrained directly by the observations, making their performance increasingly dependent on the spatial extent and information content of the available kinematics. The two approaches therefore regularize the same inverse problem in fundamentally different ways.

Which form of regularization performs better is ultimately an empirical question rather than one that can be decided from model flexibility alone. Comparisons against independent constraints show that the additional flexibility of orbit-superposition models does not automatically translate into greater accuracy. When evaluating dynamical mass models against CO gas rotation curves across 54 galaxies, \citet{Leung2018} found that JAM yielded measurement errors at $0.8$--$1.6 R_{\rm e}$ that were approximately 1.7 times smaller than those from Schwarzschild models. Tests on numerical simulations by \citet{Jin2019} similarly found a factor of 1.6 smaller scatter for JAM in recovering the total mass within $1 R_{\rm e}$, with neither study detecting a systematic bias. These results illustrate the familiar bias--variance trade-off of inverse modelling: additional model freedom is advantageous when demanded and constrained by the data, while physically motivated restrictions can improve robustness when the observations provide insufficient information to determine that freedom independently.

\subsection{Agreement between dynamical modelling methods}
\label{sec:intro_benchmarks}

The empirical comparison between JAM and Schwarzschild methodologies provides no evidence that either framework should be regarded a priori as the more accurate estimator of BH mass; rather, their different assumptions and data requirements make them complementary approaches whose relative performance depends on the information content of the observations.

\begin{figure}
    \centering
    \includegraphics[width=\columnwidth]{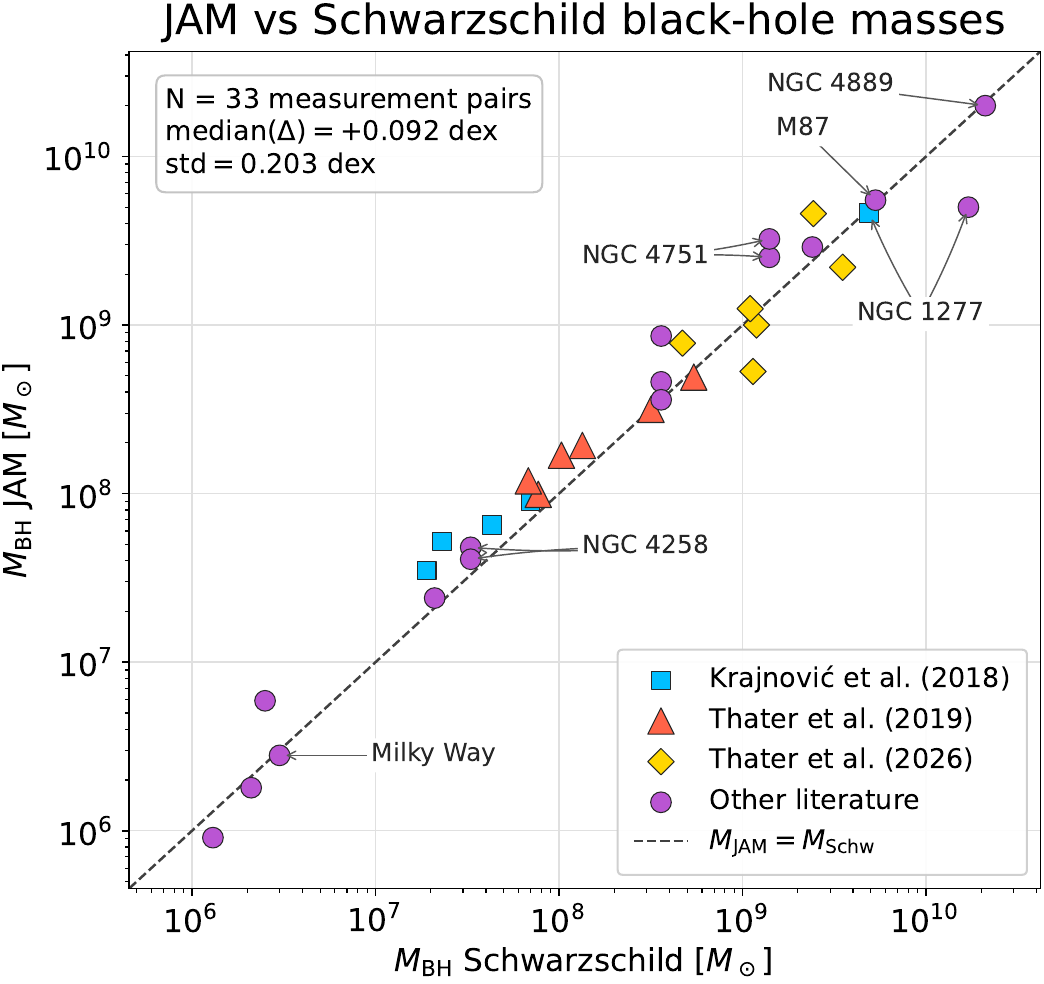}
    \caption{Comparison of supermassive black hole masses $M_\mathrm{BH}$ measured using Jeans Anisotropic Modelling (JAM) and Schwarzschild orbit-superposition models for a compilation of $N=33$ published measurement pairs. Error bars are omitted for clarity. The dashed line traces the 1:1 relation. The two methods are highly consistent, showing a median logarithmic difference $\log_{10}(M_{\rm BH,JAM}/M_{\rm BH,Sch})=+0.092$\,dex (a factor of $1.24$) and a standard deviation of $0.203$\,dex. The latter corresponds to a factor of $10^{0.203}\simeq1.60$ scatter in the JAM-to-Schwarzschild $M_\mathrm{BH}$ ratio. In the limit of equal and independent uncertainties for the two modelling techniques ($\sigma_\Delta=\sqrt{2}\,\sigma$), this implies an individual $1\sigma$ uncertainty of $0.203/\sqrt{2}=0.144$\,dex for each method, equivalent to a multiplicative factor of $1.39$ (approximately $+39/-28$ per cent). Several benchmark systems discussed in the text are labelled. Both JAM\textsubscript{cyl} and JAM\textsubscript{sph} are shown when available; when both triaxial and axisymmetric Schwarzschild models are available, the triaxial result is adopted. Upper limits are excluded. The individual measurements, modelling choices, and references used to construct this figure are listed in \autoref{tab:jam_schw_compilation}.}
    \label{fig:jam_vs_schw}
\end{figure}

To quantify this empirical consensus, \autoref{fig:jam_vs_schw} compiles $N=33$ published JAM--Schwarzschild measurement pairs for 29 distinct stellar systems. Defining $\Delta=\log_{10}(M_{\rm BH,JAM}/M_{\rm BH,Sch})$, the compilation demonstrates that the two methods yield highly consistent BH masses. There is a median offset of only $+0.092$ dex (a factor of $1.24$, or $24\%$), meaning there is no significant systematic difference between the methods compared with their object-to-object scatter of $0.203$ dex (a factor of $1.60$). If the two methods contribute equally and independently to this scatter ($\sigma_\Delta=\sqrt{2},\sigma$), the implied uncertainty of an individual BH measurement is $0.203/\sqrt{2}=0.144$ dex (a factor of $1.39$, or $39\%$). For clarity, the figure excludes the preliminary JAM--Schwarzschild comparisons presented by \citet{Cappellari2010conf}, but including them leaves this conclusion essentially unchanged: the sample increases to $N=58$ measurement pairs, the median offset decreases slightly to $+0.087$ dex (a factor of $1.22$, or $22\%$), while the scatter increases modestly to $0.266$ dex (a factor of $1.85$), likely due to the lower resolution of those early data. Thus, the agreement between the two modelling approaches is insensitive to whether these earlier preliminary measurements are included. A detailed discussion of historical comparisons, independent geometric validations, and the individual measurements used to construct the figure is provided in \autoref{app:jam_schw_compilation}.

A crucial comparison, particularly relevant for this paper, is the recent study by \citet{Thater2026}, who found consistent BH masses between triaxial DYNAMITE and JAM models for six massive galaxies showing clear evidence of triaxial shapes. Together with the independent JAM--Schwarzschild comparisons for M87 \citep{Simon2024,Liepold2023} and NGC~4889 \citep{Shetty2020,McConnell2012}, these constitute eight particularly relevant tests, as all involve massive weakly-triaxial slow-rotator early-type galaxies of the same general class as those targeted by our JWST program. NGC~4889 is especially noteworthy: it is one of the two massive BCGs of the Coma cluster and, being substantially more massive than M87, provides an even closer local analogue to the extreme galaxies in our sample. The \citet{Thater2026} study also vividly illustrated \citep[fig.~4]{Thater2026} how JAM natively provides the structural priors required to extract robust BH masses using \textit{only} the innermost high-resolution kinematics, whereas Schwarzschild models required the inclusion of large-field kinematics to reach comparable constraints.

Given that our JWST program focuses on exquisite, high-resolution nuclear kinematics but does not currently provide the large-field IFS coverage needed to fully constrain Schwarzschild models, JAM is optimally matched to the information content of the available data. The empirical agreement discussed above therefore supports our use of JAM for the present analysis. Nevertheless, because JAM and Schwarzschild methods regularize the dynamical inverse problem in fundamentally different ways, applying both approaches to the same galaxies would provide a valuable independent test of modelling systematics. We have begun acquiring the complementary large-field IFS observations required for such comparisons, with data already available for a subset of the sample. As additional observations become available, independent Schwarzschild modelling will provide an important future test of whether the BH masses inferred here remain robust to the choice of dynamical framework.

\section{Summary of Observational Data and Models}
\label{sec:data_summary}

In Paper I, we presented the full-sky sample selection, data reduction, and the foundational data products for the 12 extreme early-type galaxies ($M_\star \gtrsim 2\times10^{12}\,\mathrm{M}_\odot$) in our JWST program. To make the present dynamical analysis self-contained, we briefly summarize the construction of the photometric mass models and the extraction of the stellar kinematics. We refer the reader to Paper I for all observational details, exposure times, diagnostic plots, and tabular data.

\subsection{Multi-scale MGE Photometric Models}

Accurate dynamical BH mass measurements require a precise determination of the host galaxy's deprojected 3D stellar mass distribution, spanning from the sub-arcsecond scales of the BH sphere of influence out to the extended dark matter halo. Because the limited field of view of JWST cannot capture the vast envelopes of these giant ellipticals, we constructed composite surface brightness profiles by combining our ultra-high-resolution nuclear imaging from JWST/NIRCam (primarily utilizing the F250M filter) with wide-field ground-based $z$-band imaging from the DESI Legacy Surveys \citep{Dey2019}. For three targets lacking NIRCam coverage, we substituted the central photometry with collapsed JWST/NIRSpec flux cubes.

We parametrized the combined surface brightness profiles using the Multi-Gaussian Expansion (MGE) method \citep{Emsellem1994, Cappellari2002mge}. The space- and ground-based datasets were seamlessly stitched together using the \texttt{mge.PhotometryMatch} algorithm, which analytically solves for the optimal flux scaling of the DESI data and the residual sky background in the JWST images over a defined overlap region. The joint 1D profiles were then fitted simultaneously using regularized MGE optimization, rigorously convolving the models with the respective instrumental point spread functions (PSFs) at each spatial scale.

\subsection{Spatially Resolved Stellar Kinematics}

The spatially resolved stellar kinematics were derived from our JWST/NIRSpec integral-field spectroscopy, utilizing the high-resolution \texttt{G235H} grating. At the distances of our sample, the $3\arcsec \times 3\arcsec$ field of view optimally covers the central few kiloparsecs, penetrating well within the BH spheres of influence. To achieve the requisite signal-to-noise ratio ($\sn \approx 100$) for measuring the higher-order moments of the line-of-sight velocity distribution while preserving maximum spatial resolution in the galaxy cores, the spaxels were geometrically binned using the Centroidal Power Diagram method (\textsc{PowerBin}; \citealt{Cappellari2025}).

We extracted the stellar kinematics (velocity $V$, and velocity dispersion $\sigma$) using the penalized pixel-fitting method (\textsc{pPXF}; \citealt{Cappellari2004, Cappellari2017, Cappellari2023}). To fully exploit the continuous, telluric-free $1.66-3.17\,\mu\mathrm{m}$ spectral window provided by JWST, we modelled the stellar continuum and absorption features (most notably the deep $^{12}$CO bandheads) using the updated BOSZ synthetic stellar library \citep{Meszaros2024}. 

To reduce the redundancy of modern synthetic grids without artificially mixing templates, we introduced a novel compression approach based on near-separable Non-Negative Matrix Factorization (NMF) via the Successive Nonnegative Projection Algorithm \citep[SNPA;][]{Gillis2014, Gillis2020}. This mathematically optimal procedure identified a highly compressed, non-redundant subset of exactly 100 synthetic spectra that encapsulates the full additive spanning power of the parent BOSZ library. Using this optimal template set, the final kinematics were extracted utilizing the scale-invariant sigma-clipping implemented in \textsc{pPXF} v9.5, ensuring exquisitely clean velocity and dispersion fields uncontaminated by residual detector artifacts.

\section{JAM Dynamical Modelling}
\label{sec:jam_modelling}

In this study, we measure the central BH masses of our ultramassive galaxy sample by applying the Jeans Anisotropic Modelling  \citep[JAM][]{Cappellari2008,Cappellari2020,Cappellari2026jam}, method\footnote{\url{https://pypi.org/project/jampy/}} v9.0. Specifically, we utilize the recently introduced spectral solver for the axisymmetric Jeans equations \citep{Cappellari2026jam}, which provides immense flexibility and computational speed while allowing for completely general velocity anisotropy distributions. Below, we outline the theoretical framework, our geometrical assumptions, and the specific implementations adopted for these extreme systems.

\subsection{The Spectral Jeans Anisotropic Modelling Framework}
\label{sec:jam_framework}

The JAM method models the kinematics of a steady-state, axisymmetric stellar system by solving the Jeans equations \citep{Jeans1915}. Traditionally, this is achieved via semi-analytic quadratures that require restrictive assumptions about the orbital anisotropy. We overcome these limitations by employing the spectral JAM\textsubscript{sph} solver \citep{Cappellari2026jam}, which operates within a spherically aligned velocity ellipsoid framework. In this coordinate system $(r, \theta, \phi)$, the mixed second velocity moments vanish ($\overline{v_r v_\theta} = \overline{v_r v_\phi} = \overline{v_\theta v_\phi} = 0$). 

A key mathematical breakthrough of the spectral solver is that, rather than solving for the rapidly varying pressure term $\nu \overline{v_r^2}$, it solves directly for the intrinsic radial velocity dispersion $\overline{v_r^2}$ over a logarithmic radial grid. The system's state is entirely described by a single, linear first-order partial differential equation:
\begin{equation}
-\cos\theta \frac{\partial \overline{v_r^2}}{\partial \ln r} + (1-\beta)\sin\theta \frac{\partial \overline{v_r^2}}{\partial \theta} + \mathcal{C}(r, \theta)\, \overline{v_r^2} = r \frac{\partial\Phi(R, z)}{\partial z},
\label{eq:jam_pde}
\end{equation}
where $\Phi$ is the total gravitational potential, and $\mathcal{C}(r, \theta)$ is a coefficient that absorbs the logarithmic gradients of the tracer density $\nu$ and the angular derivatives of the anisotropy \citep[for a full derivation, see][]{Cappellari2026jam}. The anisotropy itself is defined generally as $\beta(r, \theta) \equiv 1 - \overline{v_\theta^2}/\overline{v_r^2}$. Once $\overline{v_r^2}$ is evaluated spectrally, the azimuthal second moment $\overline{v_\phi^2}$ is recovered algebraically, allowing us to project the total $V_{\rm rms} \equiv \sqrt{V^2 + \sigma^2}$ onto the sky plane to compare directly with our JWST/NIRSpec observations.

\subsection{Velocity Ellipsoid Alignment and Anisotropy}
\label{sec:jam_anisotropy}

Previous implementations of the Jeans Anisotropic Multi-Gaussian Expansion (JAM) method required choosing between a cylindrically aligned velocity ellipsoid (\textsc{JAM}\textsubscript{cyl}; \citealt{Cappellari2008}) and a spherically aligned one (\textsc{JAM}\textsubscript{sph}; \citealt{Cappellari2020}). Orbit integrations in realistic galactic potentials demonstrate that orbital envelopes fundamentally follow a radial geometry, establishing spherical alignment as the most physically robust global approximation \citep{Cappellari2026jam}. However, in flattened, fast-rotating early-type galaxies (ETGs), the velocity dispersion ratio often approximates cylindrical alignment (i.e., a nearly constant $\sigma_z/\sigma_R$).

The new spectral JAM solver resolves this tension by operating natively in spherical coordinates while flexibly allowing for angular variations in the anisotropy profile $\beta(r, \theta)$. To explore whether the assumed velocity ellipsoid orientation influences our inferences, we initially considered two distinct configurations:
\begin{enumerate}
    \item \textbf{Standard Spherical Alignment:} A globally constant anisotropy ratio in spherical coordinates, where $\sigma_\theta/\sigma_r$ (and thus $\beta$) is assumed to be spatially constant.
    \item \textbf{Approximate Cylindrical Alignment:} An angularly varying anisotropy designed to emulate cylindrical alignment along the principal axes. Following \citet[eq.~37]{Cappellari2026jam}, the ratio varies smoothly with polar angle $\theta$:
    \begin{equation}
        \sigma_\theta/\sigma_r = \mathcal{R}^{-\cos(2\theta)} \implies \beta(\theta) = 1 - \mathcal{R}^{-2\cos(2\theta)},
    \end{equation}
    where $\mathcal{R}$ parameterizes the anisotropy ratio. This formulation reproduces the global kinematics of traditional cylindrical models while circumventing the unphysical singularities that \textsc{JAM}\textsubscript{cyl} produces along the symmetry axis.
\end{enumerate}

During our initial fitting stage, the models with approximate cylindrical alignment consistently yielded clearly unacceptable fits to the observed kinematics, producing systematic residuals and failing to match the two-dimensional velocity dispersion fields. Consequently, we discarded this parametrization and did not employ cylindrical-like models in any of our final analyses.

This empirical outcome is reassuring and physically well motivated. Massive, core-like ETGs such as those in our sample are slow rotators shaped by major dry mergers and are not expected to contain fast rotating stellar disks \citep[e.g.][]{Emsellem2011p3,Cappellari2013p20,Krajnovic2018prolate,Krajnovic2026}. See reviews in \citet{Cappellari2016,Cappellari2026}. Near the central supermassive black hole, where the gravitational potential is overwhelmingly dominated by a central point mass and the surrounding round stellar core, the stellar orbits are naturally expected to sustain a spherically aligned velocity ellipsoid. Having ruled out cylindrical-like alignments on both observational and physical grounds, we exclusively adopt the standard spherically aligned solver (\textsc{JAM}\textsubscript{sph}) for all subsequent kinematic fits and black hole mass determinations.

\subsection{Galaxy Shapes: Oblate versus Prolate Geometries}
\label{sec:jam_shapes}

The extreme mass of our galaxy sample ($M_\star \gtrsim 2\times10^{12}\,\mathrm{M}_\odot$) places them in a unique structural regime. While massive slow-rotator early-type galaxies with  $2\times10^{11}\lesssim M_\star \lesssim 10^{12}\,\mathrm{M}_\odot$ are generally weakly triaxial with an oblate tendency \citep{Weijmans2014p24, Li2018shapes}, the most massive galaxies in the Universe show compelling evidence of transitioning toward strongly prolate shapes \citep{Krajnovic2018prolate, Krajnovic2026}. Historically, dynamical BH mass measurements across the literature have overwhelmingly assumed oblate axisymmetry. 

To account for this structural shift, our analysis explicitly tests both oblate and prolate extremes. The axisymmetric JAM formalism effortlessly supports prolate shapes with no changes to the underlying differential equations or deprojection machinery. Constructing a prolate model requires four simple modifications:
\begin{enumerate}
    \item We rotate the assumed symmetry axis by 90 degrees, aligning it with the photometric \textit{major} axis rather than the minor axis.
    \item For a given set of MGE parameters fitted to the photometry, the intrinsic axial ratio of each Gaussian component is inverted to $1/q_j$, where $q_j < 1$ is the observed flattening.
    \item The spatial dispersion of each Gaussian is scaled to $\sigma_j q_j$, stretching the models along the new symmetry axis.
    \item For any physically motivated priors placed on the velocity anisotropy (e.g., assuming $\sigma_z < \sigma_R$ for an oblate system), we invert the prior to $\sigma_z > \sigma_R$ to reflect the prolate geometry.
\end{enumerate}
Testing both oblate and prolate boundaries allows us to marginalize over the unknown intrinsic 3D geometry of these ultramassive systems.

\subsection{The Role of Dark Matter and Photometric Twists}
\label{sec:jam_dm_twists}

When employing Schwarzschild orbit-superposition models, one typically requires wide-field integral-field kinematics to constrain the extended orbital families. This necessitates the explicit inclusion of a dark matter halo, as the kinematics extend into radii where dark matter and stellar population gradients become non-negligible. In contrast, the JAM method can robustly recover the central potential using only the high-resolution nuclear kinematics, extending just a few times beyond the BH sphere of influence \citep{Thater2026}. Because our JWST data focuses exclusively on the central few kiloparsecs, the mass contribution of an extended dark matter halo is completely negligible, and we thus omit it from our models.

Another phenomenon commonly associated with massive triaxial galaxies is the presence of isophotal twists. Traditional wisdom suggests that such twists require fully triaxial deprojections, precluding the use of a fixed position angle (PA) for the MGE components. However, a careful analysis of our sample in Paper I reveals no genuine evidence of triaxial photometric twists. While we observe clear deviations from a fixed PA, these deviations are never point-symmetric. Instead, the isophotes exhibit asymmetric elongations with wandering centers. 

These features are classic signatures of unrelaxed tidal debris and non-equilibrium states driven by the dry mergers that assembled these galaxies. Crucially, no steady-state dynamical model---whether JAM or Schwarzschild---can accurately model non-equilibrium kinematics. Fortunately, this is largely irrelevant for our specific study. We are modelling the innermost nuclear regions where the dynamical timescales are orders of magnitude shorter than the galaxy outskirts, meaning the central regions are well-relaxed. The only exception is a slight mis-centering of the nuclei relative to the broader galaxy in some objects, likely indicative of BH sloshing following a recent binary BH merger. While fascinating from an evolutionary standpoint, the spatial scale of this off-centering is small enough that it induces only negligible systematic uncertainties in our BH mass measurements. Consequently, we model all galaxies using MGEs with a fixed PA, confident that large-scale asymmetries will not bias the nuclear dynamics.

\subsection{Anisotropy Priors}
\label{sec:jam_priors}

To break the mass--anisotropy degeneracy inherent to Jeans modelling, one can apply physically motivated priors on the orbital distribution based on empirical trends derived from large samples of galaxies \citep[see review in][section ``Dynamical modeling and orbital anisotropy'']{Cappellari2026}.

For our models utilizing the approximate cylindrical alignment (where the anisotropy varies as a function of polar angle), we found that the parameter space naturally regularized without the need for explicit constraints on the anisotropy ratio $\mathcal{R}$.

For the standard JAM\textsubscript{sph} models, in which the anisotropy is independent of polar angle, we instead allow it to vary radially. In \citet{Simon2024}, we chose, primarily for mathematical convenience, to parametrize the anisotropy in terms of the usual spherical anisotropy variable \citep[e.g.][eq.~4-53b]{Binney1987}
$$
\beta(r)=1-\frac{\sigma_\theta^2}{\sigma_r^2},
$$
while adopting a flexible logistic function of $\lg r$ to describe its radial variation. The latest spectral implementation of JAM \citep{Cappellari2026jam} has removed the computational restrictions that motivated the choice of $\beta$ as the parametrized variable, while retaining the freedom to use the same convenient logistic radial profile. We are therefore now free to choose the anisotropy variable based purely on its physical interpretation, statistical behaviour, and visual convenience, rather than on mathematical convenience.

Although $\beta$ has a familiar physical interpretation, it is not an ideal variable for parametrization or Bayesian sampling because its allowed domain, $-\infty<\beta<1$, is strongly asymmetric between radial and tangential anisotropy. This asymmetry also gives disproportionate visual weight to tangential anisotropy when plotting $\beta$, making radial and tangential deviations from isotropy appear intrinsically different even when the corresponding dispersion ratios are reciprocal. Using the dispersion ratio $\sigma_\theta/\sigma_r$ reduces this asymmetry, but does not eliminate it: radial and tangential configurations related by interchanging $\sigma_r$ and $\sigma_\theta$ correspond to reciprocal rather than opposite values of the ratio.

A particularly simple symmetric parametrization is instead provided by the logarithmic dispersion ratio
$$
\eta\equiv\lg\frac{\sigma_\theta}{\sigma_r}.
$$

Radial and tangential anisotropy are then treated on an equal footing, because the interchange $\sigma_r\leftrightarrow\sigma_\theta$ corresponds simply to $\eta\leftrightarrow-\eta$, with isotropy at $\eta=0$. This symmetry is useful both statistically and visually: equal reciprocal departures of $\sigma_\theta/\sigma_r$ from unity appear at equal distances on opposite sides of zero, without assigning greater visual weight to either radial or tangential anisotropy. Moreover, $\eta$ retains a simple interpretation close to isotropy, which is the regime empirically occupied by real galaxies. Writing $\sigma_\theta/\sigma_r=1+\epsilon$, with $|\epsilon|\ll1$, and using
$$
\ln(1+\epsilon)=\epsilon-\frac{\epsilon^2}{2}+\mathcal{O}(\epsilon^3),
$$
gives
$$
\eta=\lg(1+\epsilon)
=\frac{1}{\ln 10}\left[\epsilon-\frac{\epsilon^2}{2}+\mathcal{O}(\epsilon^3)\right]
\simeq\frac{\epsilon}{\ln 10}.
$$

Thus, to first order near isotropy, $\eta$ scales directly with the fractional deviation of $\sigma_\theta/\sigma_r$ from unity, while remaining exactly antisymmetric under interchange of the radial and tangential dispersions. These properties make $\eta$ a natural variable for describing and visualizing anisotropy and for sampling it with symmetric Bayesian priors.

For these reasons, rather than parametrizing $\beta$ as a logistic function of $\lg r$ as in \citet{Simon2024}, we adopt the same functional form for $\eta$:
$$
\eta(r)\equiv\lg\frac{\sigma_\theta}{\sigma_r}
=\eta_0+\frac{\eta_\infty-\eta_0}{1+(r_a/r)^\alpha},
$$
where $\eta_0$ and $\eta_\infty$ are the asymptotic inner and outer anisotropies, $r_a$ sets the transition radius, and $\alpha$ controls the sharpness of the transition. The corresponding conventional anisotropy is
$$
\beta(r)=1-10^{2\eta(r)}.
$$

We impose priors on this profile such that the orbits are tangentially biased within a region of order a few times the BH sphere of influence and transition to being weakly radially biased further out. This orbital structure is strongly motivated both by theoretical predictions of core scouring by supermassive BH binaries during dry mergers and by direct empirical observations \citep[see review in][section ``Dynamical modeling and orbital anisotropy'']{Cappellari2026}. Depending on whether the structural assumption is oblate or prolate (as discussed in \autoref{sec:jam_shapes}), these priors smoothly penalize extreme velocity anisotropies, ensuring that the MCMC sampler explores physically plausible regions of parameter space without unduly restricting the data-driven fit.

\section{Results of the JAM Dynamical Modelling}
\label{sec:jam_results}

\begin{figure*}
\centering
\includegraphics[width=\textwidth]{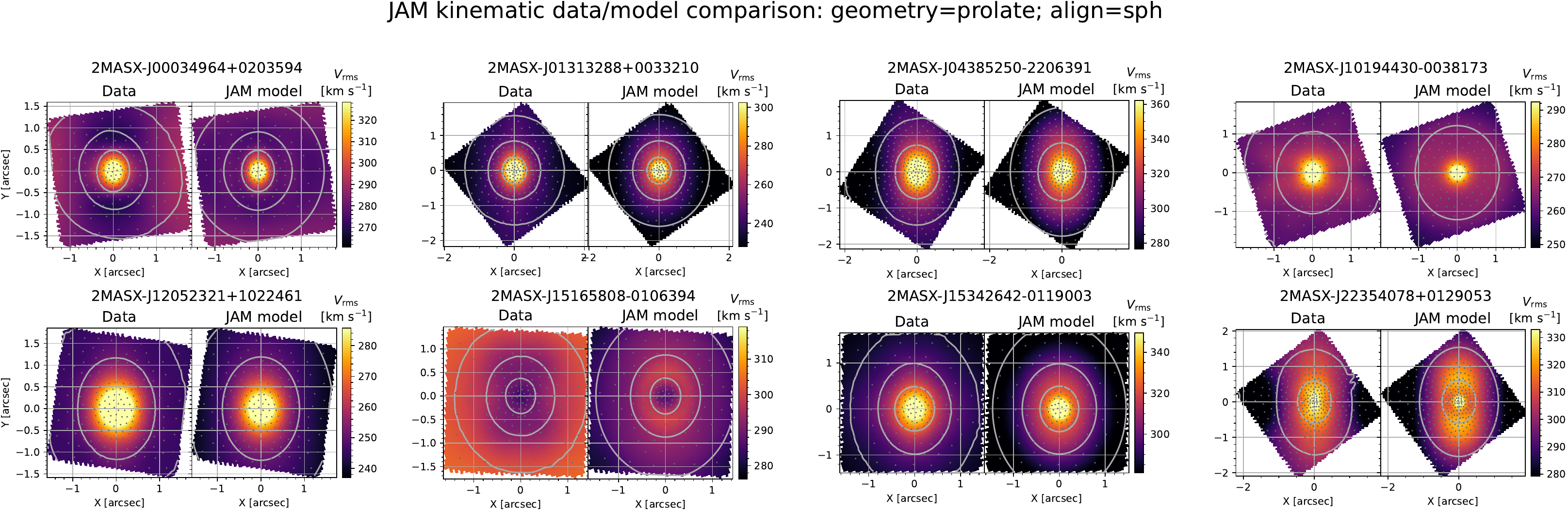}
\caption{\textbf{Spatially resolved stellar kinematics and dynamical models.} Comparison between the observed and JAM-modelled projected second velocity moment ($V_{\rm rms}=\sqrt{V^2+\sigma^2}$) for the eight ultramassive galaxies in our sample. For each galaxy (identifier above), the left map displays the even-symmetrized JWST/NIRSpec kinematic data, extracted using BOSZ $R\sim10^4$ templates, and the right map shows the corresponding best-fitting JAM model. The models assume prolate geometry and spherical alignment of the velocity ellipsoid. Grey contours indicate galaxy surface brightness in steps of 1\,mag; these are derived directly from the NIRSpec data cube and overlaid on the observed kinematics (left panels), and derived from the convolved MGE model and overlaid on the JAM model kinematics (right panels). Coordinates are relative to the galaxy centre in arcseconds, and all maps have been rotated so that the galaxy's symmetry axis aligns with the $y$-axis. Colour bars indicate $V_{\rm rms}$ in $\mathrm{km\,s^{-1}}$; colour scales are independent for each galaxy, but identical between the data and model within each pair.}
\label{fig:jam_maps_prolate}
\end{figure*}

\begin{figure*}
\centering
\includegraphics[width=\textwidth]{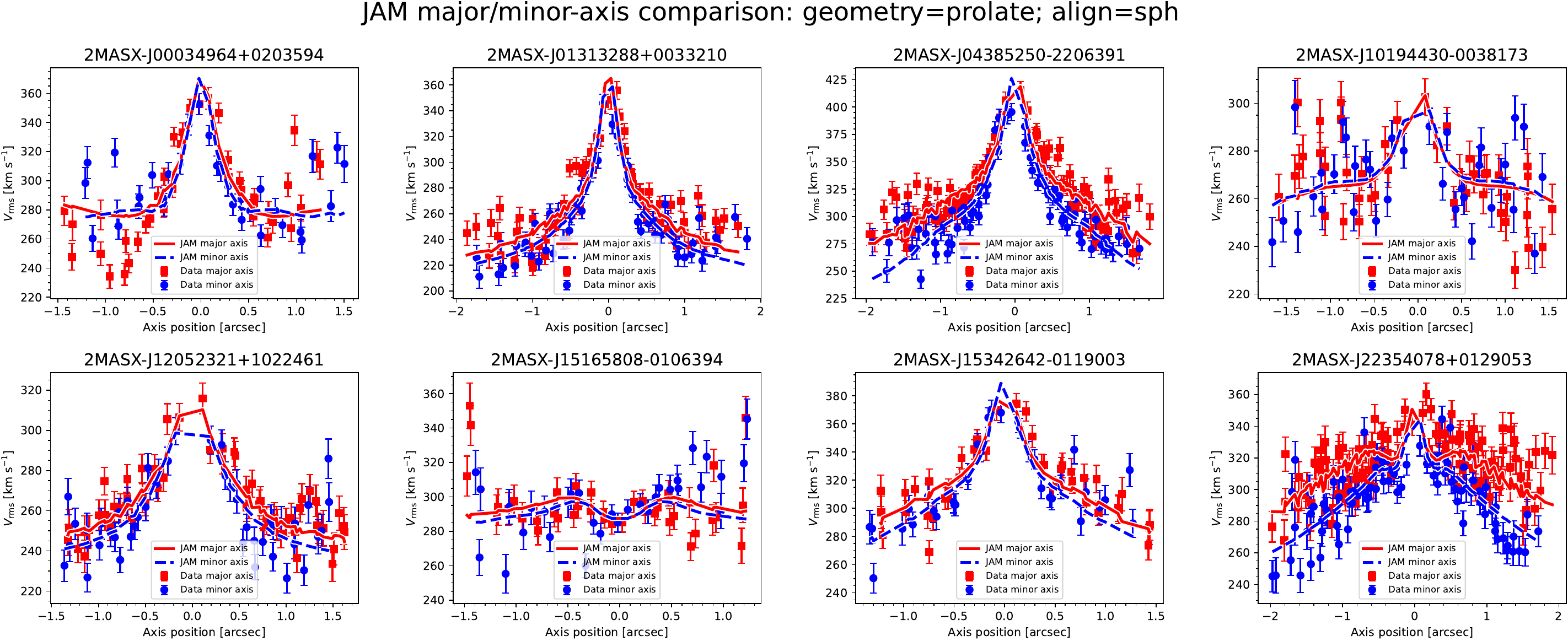}
\caption{\textbf{One-dimensional kinematic profiles and black hole signatures.} Profiles of the root-mean-square velocity ($V_{\rm rms}$) extracted along the major (red squares) and minor (blue circles) axes, comparing the JWST/NIRSpec observations to the prolate JAM\textsubscript{sph} models (red solid and blue dashed lines, respectively). Model predictions are evaluated at the median of the Bayesian posterior. As qualitatively expected for systems hosting overmassive black holes, six of the galaxies display a clear, prominent central peak in $V_{\rm rms}$, while a seventh exhibits a more modest, yet distinct, nuclear peak. Crucially, one galaxy (2MASX-J15165808-0106394) shows no evidence of a central peak. Despite the high quality of the data, its profile features a clear, small central drop; it lacks the characteristic kinematic signature of a central mass and remains entirely consistent with the complete absence of a black hole.}
\label{fig:jam_1d_prolate}
\end{figure*}

\begin{figure*}
\centering
\includegraphics[width=\textwidth]{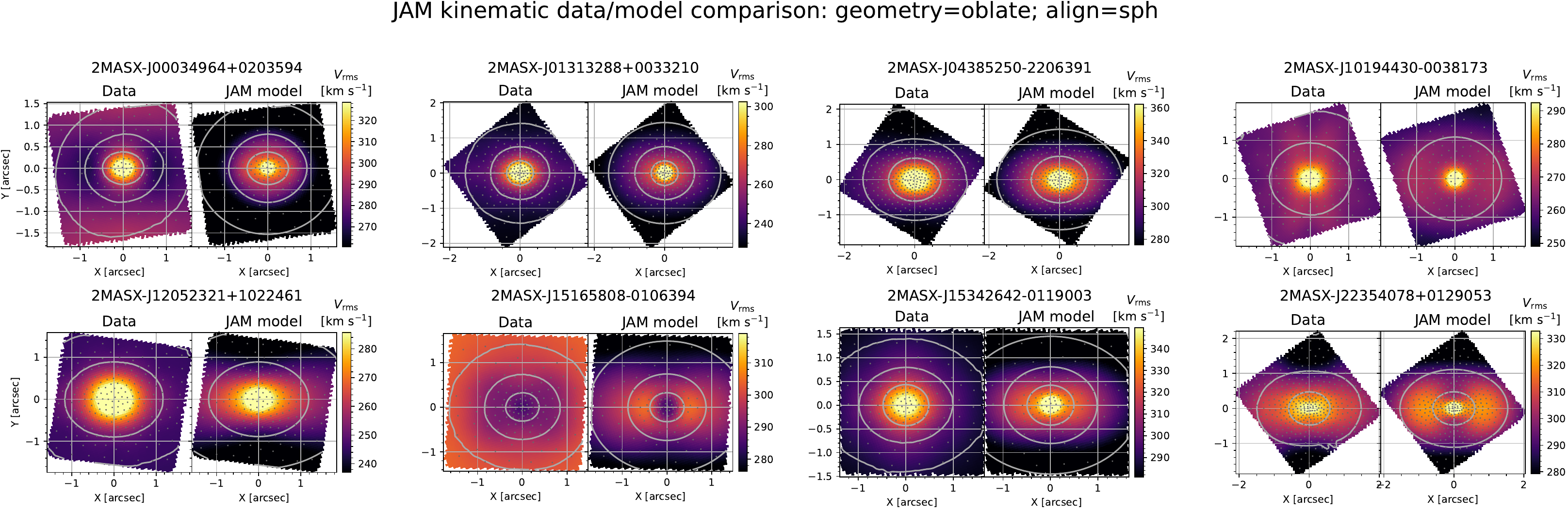}
\caption{\textbf{Two-dimensional stellar kinematics and best-fitting oblate JAM\textsubscript{sph} models.} The panels mirror those in \autoref{fig:jam_maps_prolate}, but assume an oblate rather than a prolate intrinsic geometry. The visual quality of the fits is highly comparable to the prolate scenario, highlighting that nuclear kinematics alone cannot uniquely break the degeneracy in the intrinsic three-dimensional shape. Nevertheless, although the overall differences are small, the prolate models provide a better description of the data in several cases.}
\label{fig:jam_maps_oblate}
\end{figure*}

\begin{figure*}
\centering
\includegraphics[width=\textwidth]{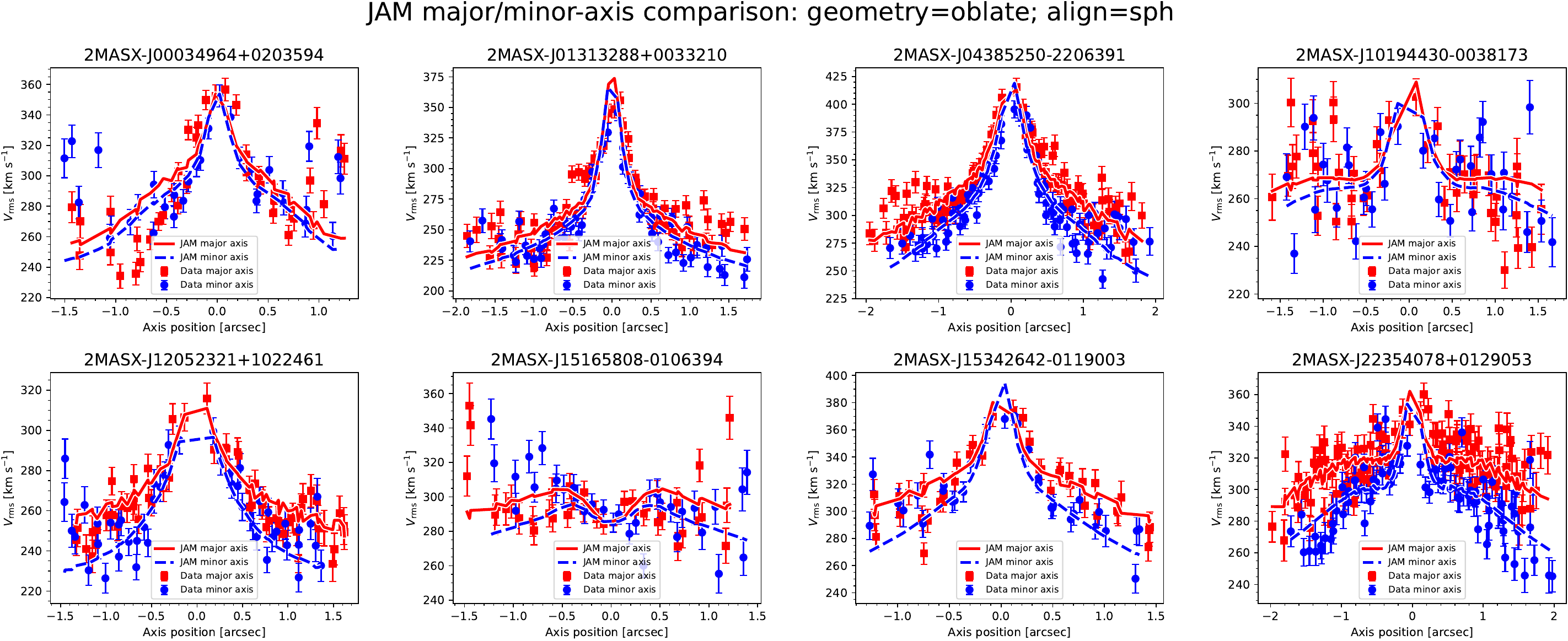}
\caption{Same as \autoref{fig:jam_1d_prolate}, but for the oblate JAM\textsubscript{sph} models shown in \autoref{fig:jam_maps_oblate}.}
\label{fig:jam_1d_oblate}
\end{figure*}

\begin{figure*}
\centering
\includegraphics[width=0.45\textwidth]{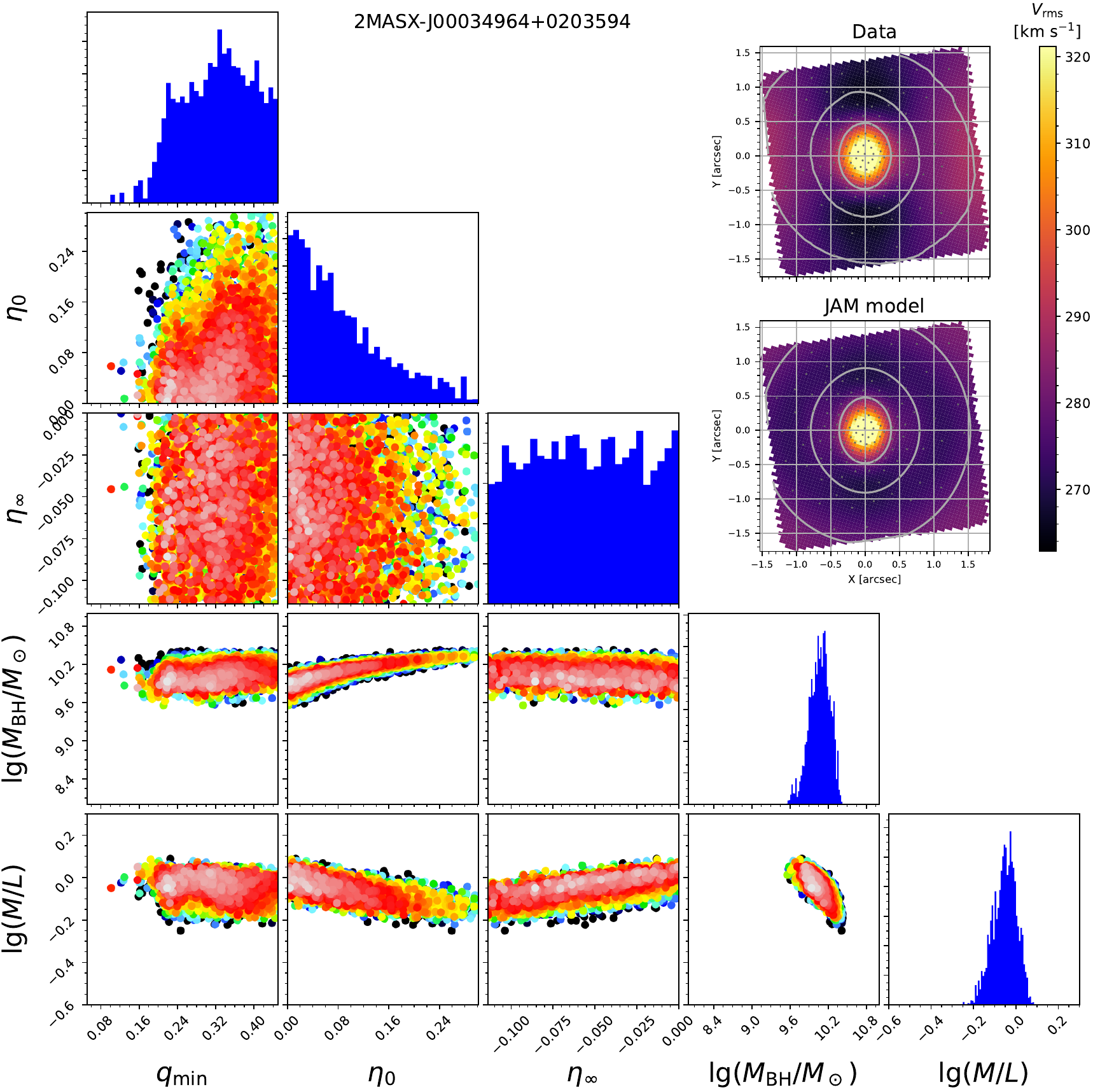}
\includegraphics[width=0.45\textwidth]{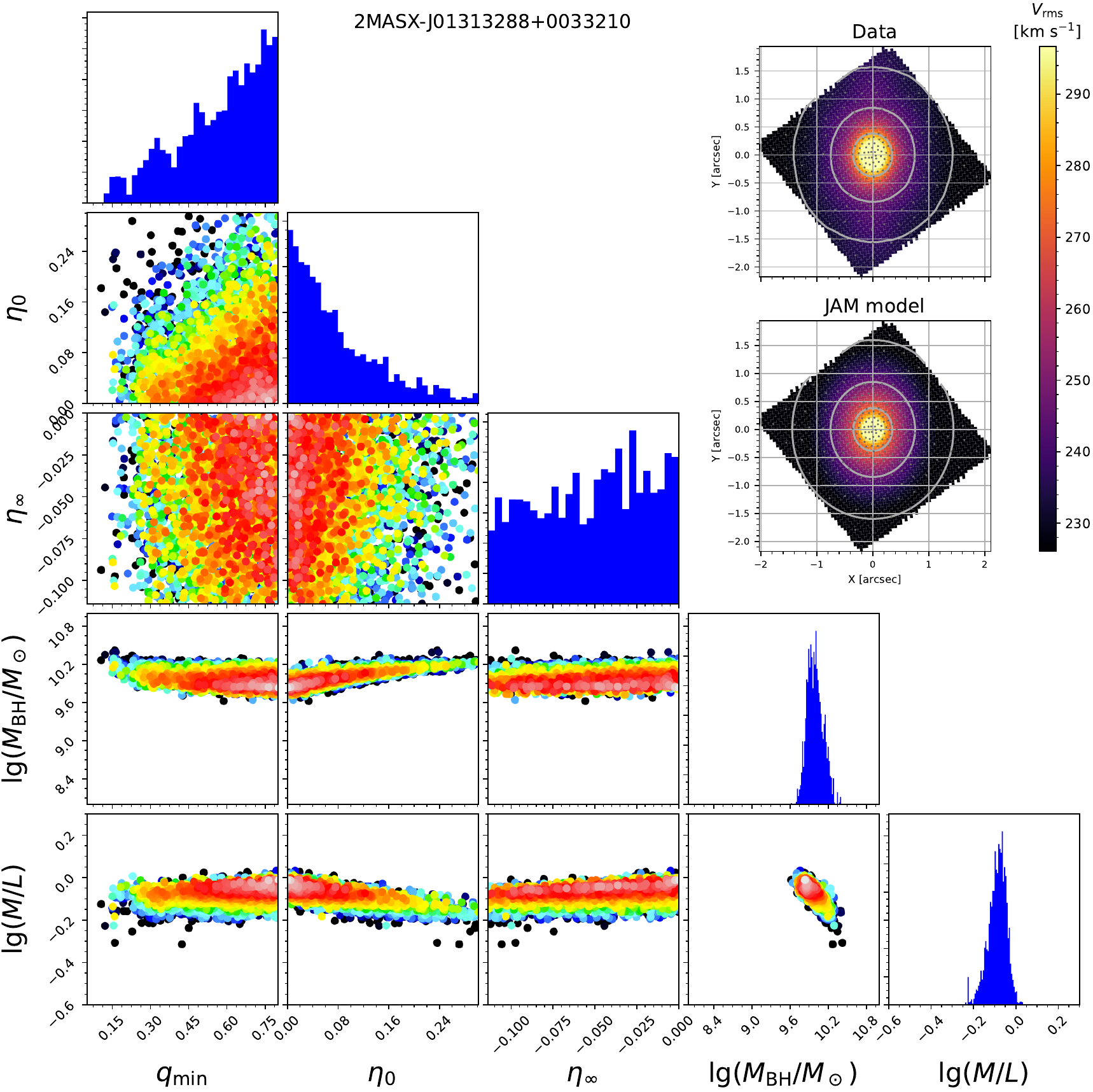}
\includegraphics[width=0.45\textwidth]{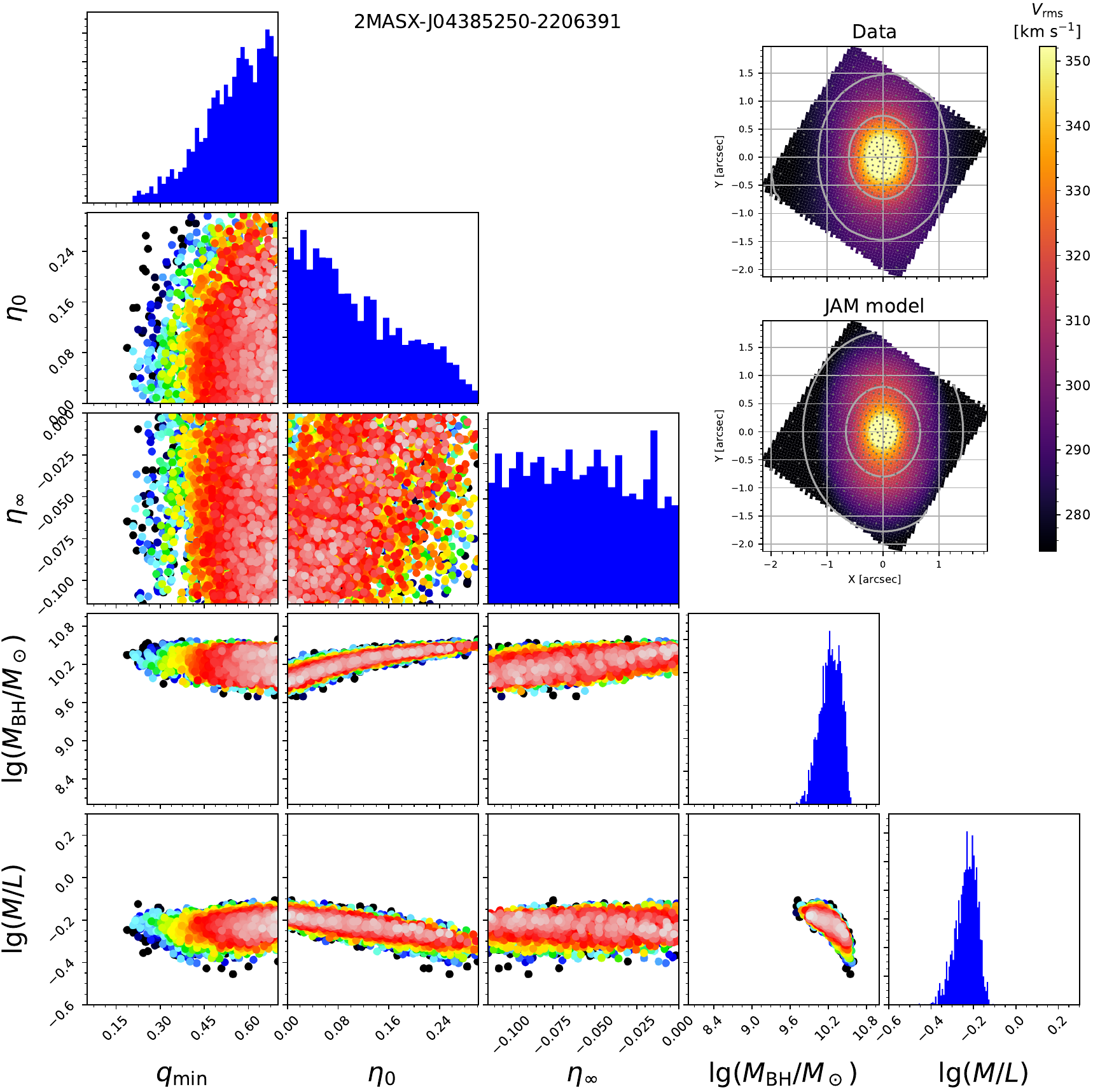}
\includegraphics[width=0.45\textwidth]{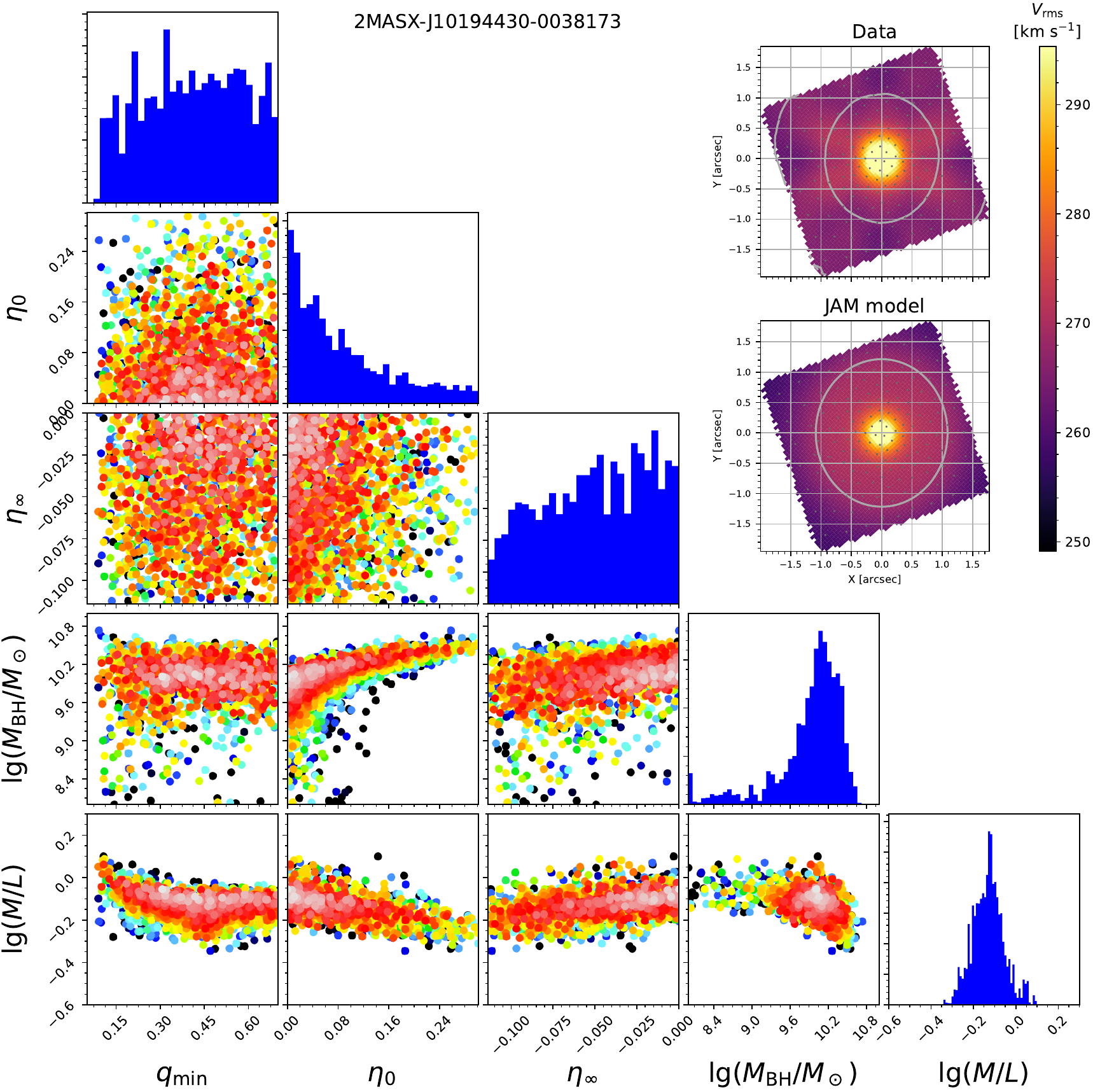}
\caption{Marginalized Bayesian posterior distributions for the prolate JAM\textsubscript{sph} models, representing the most physically plausible geometries for these ultramassive galaxies. The diagonal panels show the 1D marginalized distributions for the intrinsic-shape/inclination parameter $q_{\min}$, the inner and outer anisotropies ($\eta_0$ and $\eta_\infty$), the BH mass ($M_{\rm BH}$), and the stellar mass-to-light ratio ($M/L$). Off-diagonal panels display the 2D covariances, while the insets compare the observed and representative model $V_{\rm rms}$ maps. Crucially, as detailed in \autoref{sec:jam_posteriors}, the MCMC sampling incorporates a heuristic scaling of the input kinematic uncertainties (error inflation) by a factor of $(2 N_{\rm values})^{1/4}$. This adjustment prevents the large number of spatial bins from artificially restricting the sampler, allowing it to successfully map the full breadth of the underlying physical degeneracies shown here. Because the spatial coverage is limited to the nuclear region, the outer anisotropy $\eta_\infty$ is largely prior-dominated. The inner anisotropy $\eta_0$ generally favours near-isotropy but allows for substantial tangential configurations, which correlate directly with higher $M_{\rm BH}$. Conversely, the dependence of $M_{\rm BH}$ on the inclination parameter $q_{\min}$ is minimal.}
\label{fig:jam_corner_1}
\end{figure*}

\begin{figure*}
\centering
\includegraphics[width=0.45\textwidth]{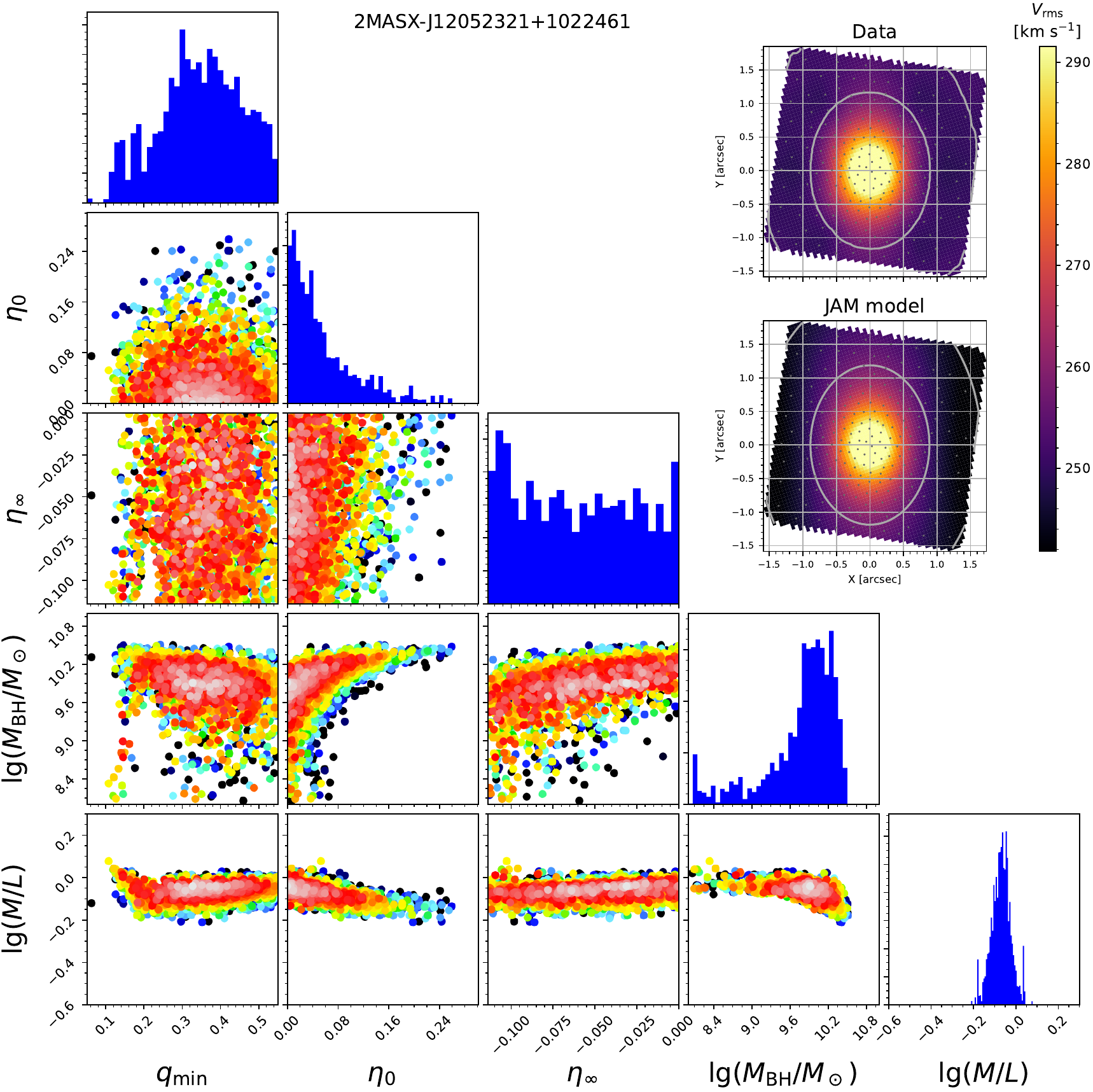}
\includegraphics[width=0.45\textwidth]{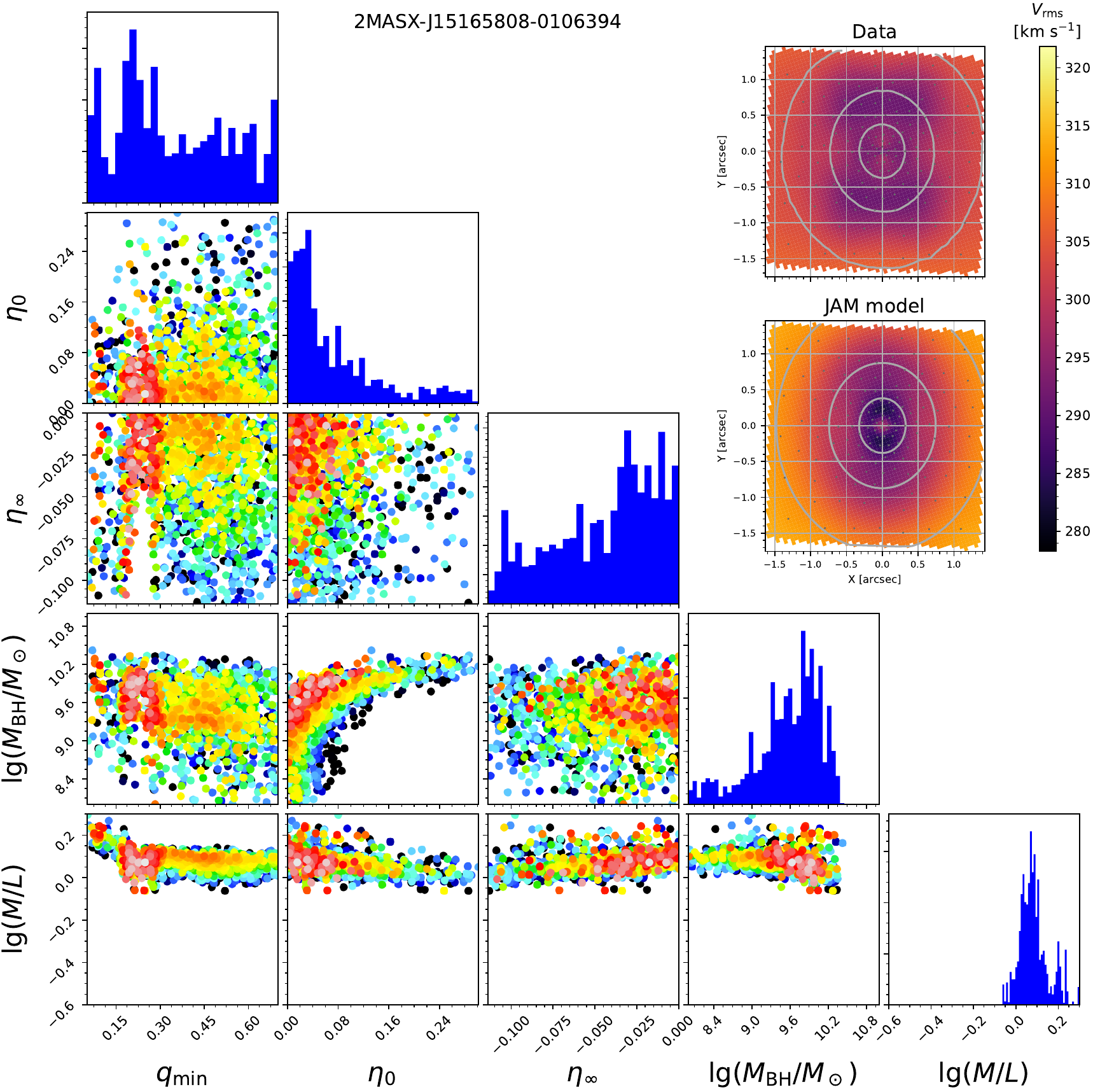}
\includegraphics[width=0.45\textwidth]{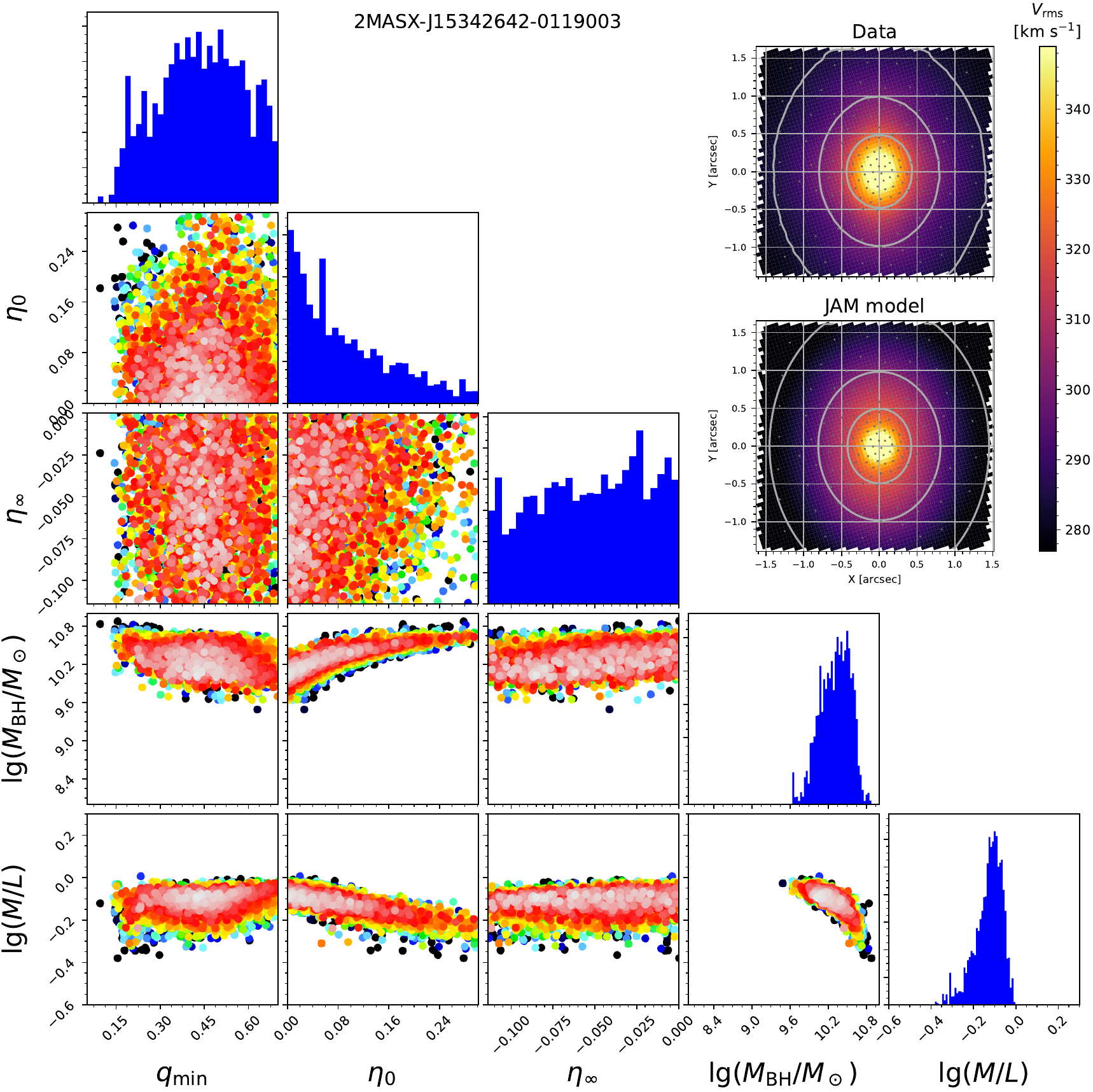}
\includegraphics[width=0.45\textwidth]{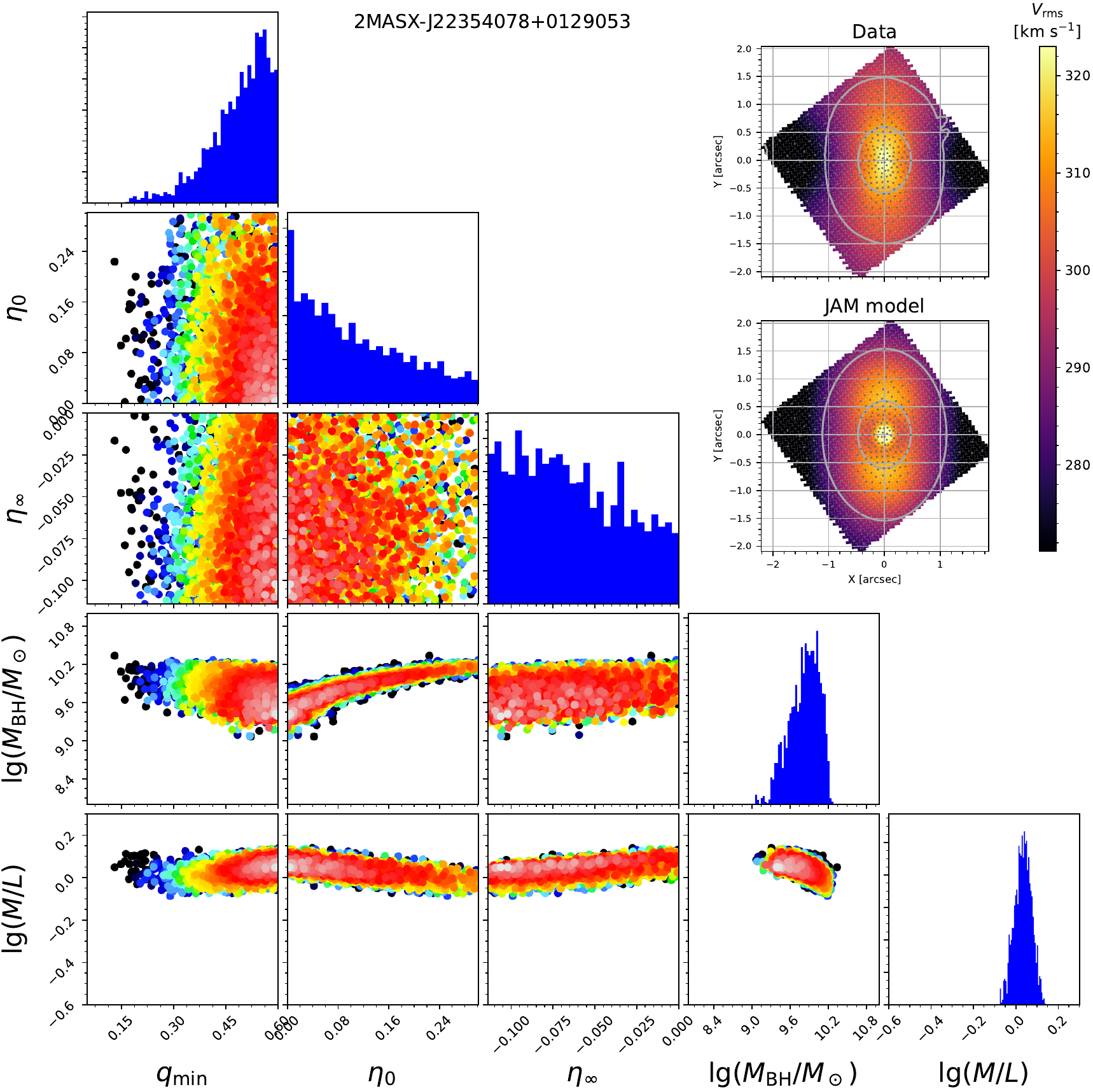}
\caption{Same as \autoref{fig:jam_corner_1}, for the remaining galaxies in the sample.}
\label{fig:jam_corner_2}
\end{figure*}

\subsection{Kinematic Fits and Intrinsic Geometries}
\label{sec:jam_fits}

To determine the stellar dynamics and central black hole masses, we optimize our JAM models utilizing a two-step framework. First, to rapidly explore the parameter space and verify that our solutions are not trapped in local minima of the likelihood surface, we employ the Levenberg--Marquardt least-squares optimization provided by the \textsc{CapFit} package \citep{Cappellari2023}. Following this initial exploration, we rigorously quantify the parameter uncertainties and their covariances by sampling the full Bayesian posterior distributions using the Adaptive Metropolis \citep{Haario2001} implemented\footnote{\url{https://pypi.org/project/adamet/}} in \textsc{AdaMet} \citep{Cappellari2013p15}. Because the two approaches yield highly consistent best-fitting parameters, we exclusively report the final parameter estimates and uncertainties derived from the Bayesian posteriors throughout the remainder of this paper.

In addition to the velocity anisotropy priors discussed in \autoref{sec:jam_priors}, we must specify our structural priors regarding the alignment of the velocity ellipsoid. The ultramassive slow rotators comprising our sample exhibit remarkably round central morphologies and become elongated only at larger radii \citep[e.g.][]{Krajnovic2018prolate}. In such systems, where the overall structure is driven by the conservation of orbital angular momentum during major dry mergers rather than by ordered rotation, a cylindrically aligned velocity ellipsoid is physically unmotivated. We thus adopt the spherically aligned models (JAM\textsubscript{sph}) as our fiducial framework, having verified that the inferred BH masses do not depend strongly on this choice.

Furthermore, as outlined in \autoref{sec:jam_shapes}, we assess the impact of the intrinsic three-dimensional geometry by running models under both oblate and prolate assumptions. While a prolate geometry is heavily favoured for extreme-mass galaxies from a structural and evolutionary standpoint \citep{Krajnovic2018prolate, Krajnovic2026}, an oblate configuration cannot be completely ruled out a priori using nuclear kinematics alone. We therefore derive kinematic fits for both geometries and ultimately combine their posterior distributions. This approach safely marginalizes over the unknown intrinsic 3D shape, incorporating the geometrical ambiguity directly into our final error budgets. For visual clarity, however, the posterior corner plots presented below (\autoref{fig:jam_corner_1} and \autoref{fig:jam_corner_2}) reflect only the prolate JAM\textsubscript{sph} models, which we consider the most physically plausible representation of our sample.

While the original, un-symmetrized two-dimensional stellar kinematics were extensively detailed in Paper I, Figures \ref{fig:jam_maps_prolate} and \ref{fig:jam_maps_oblate} present the two-dimensional kinematic maps and corresponding major- and minor-axis profiles for the prolate and oblate JAM\textsubscript{sph} models utilizing symmetrized and smoothed data visualizations. As is customary in dynamical analyses, displaying the bi-symmetrized kinematics makes it easier to perform a visual ``fit-by-eye'' evaluation, given that the underlying model is strictly axisymmetric and therefore inherently only ``sees'' the symmetric component of the data. Small asymmetric deviations in the observations have negligible effects on the global $\chi^2$ minimum of the models.

To generate these refined visual representations, we applied the spatial interpolation and smoothing procedures implemented in the \texttt{display\_bins} routine within version 3.2 of the Python package\footnote{\url{https://pypi.org/project/plotbin/}} \textsc{plotbin}. The algorithm first computes the flux-weighted barycenter of each Voronoi bin and mirrors the measured values across the galaxy's principal axes, enforcing odd parity for the mean velocity $V$ and even parity for the velocity dispersion $\sigma$ and $V_{\rm rms}$. The field is then evaluated at all original pixel locations using a linear Radial Basis Function (RBF) interpolator, implemented via the \textsc{SciPy} routine \href{https://docs.scipy.org/doc/scipy/reference/generated/scipy.interpolate.RBFInterpolator.html}{\texttt{scipy.interpolate.RBFInterpolator}}. Rather than applying a blanket smoothing, the routine employs a statistical heuristic that scales the RBF smoothing weights locally based on the noise-to-signal variance ratio. Crucially, to avoid artificially washing out the very BH signatures we aim to measure, the central-most bins are explicitly excluded from this smoothing (enforcing exact interpolation), thereby perfectly preserving the sharp central kinematic gradients.

While both geometric assumptions generally provide satisfactory descriptions of the observed kinematics, several galaxies in our sample exhibit a clear preference for prolate models. This preference is encouraging given our structural and evolutionary expectations that galaxies at this extreme mass scale transition toward prolate shapes. Nonetheless, because nuclear kinematics alone cannot unambiguously discriminate between oblate and prolate intrinsic shapes across the entire sample, treating the intrinsic geometry as a marginalized systematic uncertainty remains both robust and well justified.

When describing the model results across the sample, it is critical to emphasize that not all galaxies show evidence for ultramassive black holes. Although this paper does not directly discuss scaling relation models (which are reserved for Paper~III, \citealt{Cappellari2026p3}), it is well known that the classic dynamical signature of a supermassive black hole is a steep rise in the nuclear stellar velocity dispersion, which is expected to generally approximate a Keplerian rise $\sigma^2 \propto 1/r$ within the sphere of influence \citep{Tremaine1994}. Our very high-quality JWST kinematics indeed clearly reveal this sharp rise for five of the galaxies in our sample: 2MASX-J00034964+0203594, 2MASX-J04385250-2206391, 2MASX-J10194430-0038173, 2MASX-J12052321+1022461, and 2MASX-J15342642-0119003.

In stark contrast, a key exception is 2MASX-J15165808-0106394, which instead exhibits a distinct central drop in $V_{\rm rms}$ and $\sigma$. An intermediate case is provided by 2MASX-J22354078+0129053, which displays only a minor central $\sigma$ peak rather than an abrupt Keplerian spike. It is reassuring, as we will demonstrate in Paper~III, that these qualitative expectations and visual differences in the kinematic data indeed translate, as expected, into corresponding differences in the deviations of these galaxies from canonical black hole scaling relations.

Unfortunately, our small galaxy sample is too small to provide a comprehensive statistical picture of the demographic distribution. Nonetheless, these few striking differences already suggest that black hole accretion and growth in ultramassive galaxies may be substantially more stochastic than canonically expected. One may speculate that in some galaxies, in very dense environments, triplets of supermassive black holes may end up forming during successive dissipationless mergers, rather than the generally assumed isolated binaries \citep{Merritt2005}. In chaotic three-body systems, the least massive object tends to be the one ejected from the system via gravitational slingshot interactions \citep{Hoffman2007}, producing a less massive central supermassive black hole in the resulting merger remnant ultramassive galaxy. Indeed, tentative observational evidence for such ejected supermassive black holes wandering alone in space has already been discovered \citep{vanDokkum2026}.

Another compelling possibility is gravitational radiation recoil (kick) imparted to the coalesced black hole by anisotropic gravitational-wave emission during a binary merger \citep[e.g.][]{Campanelli2007, Blecha2016, Komossa2012}. In an ultramassive elliptical galaxy with an enormous central potential well, the recoil velocity is generally unlikely to exceed the galaxy's escape velocity ($v_{\rm esc} \gtrsim 1000-2000\,\mathrm{km\,s^{-1}}$), meaning the black hole is not completely ejected from the host system. However, even a sub-escape kick can easily displace the supermassive black hole from the galactic centre onto a large-scale, long-lived wandering orbit \citet{Gualandris2008}. Because the dynamical friction and orbital damping timescales in low-density cores can exceed hundreds of megayears, the black hole may simply not reside at the present epoch within the compact $3\arcsec \times 3\arcsec$ field of view of the NIRSpec IFU, naturally producing the observed flat or dropping nuclear velocity dispersion profiles.

\subsection{Posterior Distributions and Dynamical Degeneracies}
\label{sec:jam_posteriors}

The full Bayesian posterior distributions for the prolate JAM\textsubscript{sph} models are shown in \autoref{fig:jam_corner_1} and \autoref{fig:jam_corner_2}. These distributions are highly instructive, as they cleanly separate the parameters directly constrained by the nuclear data from those shaped by classical dynamical degeneracies. As expected from the mass--anisotropy degeneracy \citep[e.g.,][]{Binney1982, Gerhard1993}, the velocity anisotropy parameters exhibit some of the broadest distributions. The outer anisotropy $\eta_\infty$ is largely unconstrained by the data, remaining nearly flat across much of its permitted range and effectively governed by our adopted prior. This is a direct consequence of our observational strategy: the JWST/NIRSpec data purposefully target the inner few kiloparsecs to resolve the BH sphere of influence, and thus inherently lack the large radial baseline required to pin down the global orbital structure.

The inner anisotropy $\eta_0$, conversely, is better constrained. While the posteriors generally cluster near isotropy ($\eta_0 \approx 0$), they remain broad enough to permit significant tangential anisotropy. The allowed range reflects a balance between the observational constraints and the physically motivated priors discussed in \autoref{sec:jam_priors}. Importantly, the covariance between $\eta_0$ and $M_{\rm BH}$ visible in the corner plots has a clear dynamical origin. For a fixed gravitational potential, a more tangentially biased orbital distribution (lower radial velocity dispersion) places a larger fraction of the stellar kinetic energy into motions perpendicular to the radial direction. Near the projected centre, where the radial coordinate aligns primarily with the line of sight, these tangential orbits contribute less effectively to the observed line-of-sight velocity dispersion. Consequently, an elevated central mass is required to reproduce the same observed central $V_{\rm rms}$ peak, producing the positive correlation between tangential anisotropy and $M_{\rm BH}$ \citep[e.g.][fig.~10]{Simon2024}. 

The posteriors also exhibit a prominent diagonal anticorrelation between $M_{\rm BH}$ and the stellar $M/L$. This is another standard degeneracy in stellar-dynamical BH measurements: increasing the $M/L$ raises the stellar contribution to the central potential, thereby requiring a smaller supermassive BH to fit the same central kinematic peak \citep[e.g.,][fig.~6]{vanderMarel1998}. 

In contrast, $M_{\rm BH}$ exhibits minimal covariance with the intrinsic shape/inclination parameter $q_{\min}$ for the majority of the sample as generally expected \citep[e.g.][fig.6]{Verolme2002}. The near-spherical central morphology of these galaxies ensures that their projected dynamics are largely insensitive to the viewing angle; an exactly spherical model would, by construction, be completely insensitive to inclination. The relatively weak dependence of $M_{\rm BH}$ on $q_{\min}$ empirically confirms that uncertainties in inclination do not act as a major source of systematic error in our mass estimates.

Taken together, the posterior distributions highlight a fundamental strength of our Bayesian formulation. Rather than artificially fixing nuisance parameters---such as the anisotropy, stellar $M/L$, and inclination---to single externally assumed values, the MCMC framework formally marginalizes over their physically permissible ranges, exposing their full covariances. 

However, extracting meaningful statistical uncertainties from dynamical models fitted to spatially resolved datasets presents a well-documented challenge. When working with large data arrays, such as integral-field stellar kinematics or CO gas emission cubes, the sheer number of spatial bins drives the formal statistical uncertainties on the model parameters to unrealistically small values. This effectively pins the MCMC walker to a highly localized region of the posterior, masking the true breadth of the parameter space. To mitigate this issue and crudely estimate the effect of unmodelled systematic errors, we adopt a widely used variance-scaling heuristic. The necessity of this scaling was first highlighted by \citet{vandenBosch2009}, who proposed adjusting the $\Delta\chi^2$ threshold for a given confidence interval by the $1\sigma$ scatter expected in the $\chi^2$ distribution itself ($\sqrt{2 N_{\rm dof}} \approx \sqrt{2 N_{\rm values}}$). \citet{Mitzkus2017} subsequently adapted this concept for Bayesian frameworks, demonstrating that the adjustment can be seamlessly implemented by scaling the input kinematic measurement uncertainties by a factor of $(2 N_{\rm values})^{1/4}$. Although this adjustment lacks a rigorous statistical foundation, it serves as a highly practical approximation of the systematic error budget. In fact, the resulting expanded confidence intervals generally approximate the errors estimated via more robust, computationally expensive techniques, such as bootstrapping of the kinematic data. Consequently, this heuristic is widely employed to yield more realistic parameter distributions in both stellar-dynamical \citep[e.g.,][]{Mitzkus2017, Thater2019, Thater2026, Nguyen2026maser, Nguyen2026m81} and gas-dynamical \citep[e.g.,][]{Smith2019, Smith2021, Zhang2025, Dominiak2025, Ngo2025, Nguyen2020, Nguyen2021, Nguyen2022, Nguyen2026_315, Nguyen2026_4061} modelling.

Beyond yielding more realistic confidence intervals, inflating the input uncertainties allows the MCMC sampler to explore a substantially larger volume of the parameter space. This freedom is crucial for revealing possible multiple modes, broad structural degeneracies, and parameter covariances that would otherwise remain hidden. Nevertheless, because this scaling is ultimately heuristic, the resulting posterior widths cannot be strictly interpreted as formal statistical uncertainties; the bounds shown in the posterior plots should therefore be viewed as illustrative of the degeneracies rather than rigorous limits.

\subsection{Systematic Uncertainty Budget}
\label{sec:systematics}

Having established the kinematic modeling framework and explored the degeneracies inherent to the intrinsic galaxy shapes, we formalize our total uncertainty budget on $\lg M_{\mathrm{BH}}$. We do not quote or propagate the formal statistical uncertainties derived from our Bayesian posterior sampling. Given the high signal-to-noise ratio of our data, the formal statistical errors are extremely small (typically $\lesssim 0.02\ \mathrm{dex}$) and are essentially meaningless, as they implicitly assume that the chosen dynamical model and galaxy geometry are exact. In reality, the error budget is entirely dominated by systematic assumptions regarding intrinsic shape, steady state, methodological differences, and data systematics. We therefore discard the negligible statistical errors and construct our uncertainty budget purely from the dominant systematic contributors.

First, to account for the structural ambiguity discussed in \autoref{sec:jam_fits}, we define a shape-induced systematic uncertainty directly from the discrepancy between the prolate and oblate solutions:
\begin{equation}
    \sigma_{\mathrm{shape}} = \frac{1}{2} \left| \lg M_\mathrm{BH}^\mathrm{prolate} - \lg M_\mathrm{BH}^\mathrm{oblate} \right|.
\end{equation}
Because the prolate configuration is physically favored for this extreme-mass regime but an oblate geometry remains kinematically admissible, we adopt the mean of the two logarithmic mass estimates as our fiducial value, with $\sigma_{\mathrm{shape}}$ capturing the spread introduced by the unknown intrinsic deprojection geometry.

Second, in addition to geometrical assumptions internal to our dynamical modeling, external discrepancies often arise across different dynamical methods. To account for this, we incorporate a cross-method systematic uncertainty of $\sigma_{\mathrm{method}} = 0.144\ \mathrm{dex}$, derived from galaxies with multiple independent black hole mass measurements as presented in \autoref{fig:jam_vs_schw}.

Assuming that the intrinsic deprojection geometry and the broader cross-method systematics represent independent sources of error, we combine them in quadrature to define our total adopted mass uncertainty:
\begin{equation}
    \sigma_{\lg M} \equiv \sigma_{\mathrm{sys}} = \sqrt{\sigma_{\mathrm{shape}}^2 + \sigma_{\mathrm{method}}^2}.
\end{equation}
Unless explicitly stated otherwise, it is this systematic uncertainty $\sigma_{\lg M}$ that is reported in our summary tables and propagated into the scaling relation models presented in Paper~III.

\section{Conclusions}
\label{sec:conclusions}

In this second paper of the JWST Ultramassive Galaxy Sample series, we have presented robust central supermassive black hole (BH) mass measurements for 8 extreme early-type galaxies ($M_\star > 2\times10^{12}\,\mathrm{M}_\odot$). Building upon the high-fidelity Multi-Gaussian Expansion (MGE) surface brightness profiles and the exquisite, spatially resolved JWST/NIRSpec stellar kinematics introduced in Paper I, we applied the spectral Jeans Anisotropic Modelling (JAM) framework to reconstruct the nuclear dynamics of these extreme systems. 

Given the unique structural properties of galaxies residing at the very top of the mass hierarchy, our analysis explicitly tested both oblate and prolate intrinsic geometries utilizing spherically aligned velocity ellipsoids (JAM\textsubscript{sph}). By employing a comprehensive Bayesian MCMC optimization strategy, we successfully mapped the posterior distributions of the dynamical parameters. This approach allowed us to extract reliable BH masses by naturally marginalizing over classical dynamical degeneracies involving orbital velocity anisotropy, stellar mass-to-light ratios, and viewing inclination. We demonstrated that the high-resolution nuclear kinematics provided by JWST are highly effective at constraining the central BH mass within the JAM framework, independent of the unconstrained global halo properties.

Importantly, our dynamical modelling reveals that not all ultramassive galaxies host ultramassive black holes. While the classic signature of a central supermassive black hole is a steep, Keplerian-like rise in nuclear velocity dispersion ($\sigma^2 \propto 1/r$; \citealt{Tremaine1994})---a signature clearly confirmed by our JWST data in five of the galaxies---notable exceptions exist. In particular, one ultramassive galaxy exhibits an unambiguous central drop in $V_{\rm rms}$ and $\sigma$, while another displays only an intermediate, modest peak. As will be explored in Paper~III, these qualitative kinematic differences translate directly into distinct offsets from canonical BH scaling relations. Although our sample size is modest, this diversity suggests that black hole growth in extreme dry-merger regimes may be significantly more stochastic than standard evolutionary models assume. One compelling scenario is the formation of supermassive black hole triplets during sequential mergers, where chaotic three-body interactions eject the least massive black hole \citep[consistent with recent evidence for runaway black holes;][]{vanDokkum2026}, leaving a less massive remnant in the host galaxy. Another plausible mechanism is gravitational radiation recoil following a binary coalescence \citep[e.g.][]{Campanelli2007, Blecha2016, Komossa2012}; while the kick velocity is generally insufficient to exceed the deep escape velocity of an ultramassive galaxy, it can displace the black hole onto a long-lived wandering orbit outside the central $3\arcsec \times 3\arcsec$ NIRSpec field of view.

Crucially, this work serves as the vital dynamical modelling bridge within our broader observational program. While Paper I established the foundational data products and this current paper (Paper II) provides the rigorous derivation of the supermassive BH masses, the astrophysical implications of these measurements will be the focus of the next paper in this series. In Paper III, we utilize the BH masses derived here to investigate the BH--host galaxy scaling relations at the extreme high-mass end of the galaxy population. By populating this historically sparsely sampled regime, Paper III directly tests the hypothesis that the dissipationless, dry mergers responsible for assembling ultramassive galaxies fundamentally alter their co-evolutionary scaling laws, potentially driving a transition where $M_{\rm BH}$ correlates more directly with the total stellar mass rather than the stellar velocity dispersion.

%%%%%%%%%%%%%%%%%%%%%%%%%%%%%%%%%%%%%%%%%%

\section*{Acknowledgements}

This work is based on observations made with the NASA/ESA/CSA James Webb Space Telescope. The data were obtained from the Mikulski Archive for Space Telescopes at the Space Telescope Science Institute, which is operated by the Association of Universities for Research in Astronomy, Inc., under NASA contract NAS 5-03127 for JWST. These observations are associated with JWST programme 8217.   
D.D.N. acknowledges support from the ELT postdoctoral research fellowship at the Department of Astronomy, University of Michigan.

The authors used generative AI assistants to improve workflow efficiency, including language editing and programming support; all scientific ideas, analyses, interpretations, and conclusions are their own, and the authors take full responsibility for the content of this work.

 \section*{Data Availability}

All observational data, extracted kinematics, and derived data products underlying this article will be tabulated in the paper and made available as supplementary material alongside the published version. The software packages used for the dynamical modelling and statistical analysis in this work are open-source and publicly available via the Python Package Index (PyPI). The spectral Jeans Anisotropic Modelling (JAM) software \citep{Cappellari2026jam} is available at \url{https://pypi.org/project/jampy/}. The least-squares optimization package \textsc{CapFit} \citep{Cappellari2023} and the Adaptive Metropolis MCMC sampler \textsc{AdaMet} \citep{Cappellari2013p15} are available at \url{https://pypi.org/project/capfit/} and \url{https://pypi.org/project/adamet/}, respectively.

\appendix

\section{Extended Validation and Compilation of JAM and Schwarzschild Black Hole Measurements}
\label{app:jam_schw_compilation}

\subsection{From two-integral Jeans models to modern JAM} 

Historically, early empirical comparisons may have contributed to a perceived hierarchy between modelling techniques. For instance, \citet[fig.~1.2]{Gebhardt2004} showed that the two-integral $f(E,L_z)$, semi-isotropic ($\sigma_R=\sigma_z$; $\langle v_Rv_z\rangle=0$) Jeans models of \citet{Magorrian1998}, based on low-resolution ground-based kinematics, produced BH masses that were systematically larger than fully general Schwarzschild values derived from high-resolution HST kinematics. While this historical episode provided an empirical reason for caution when using restrictive two-integral models, modern Jeans Anisotropic Modelling \citep[JAM;][]{Cappellari2008,Cappellari2020,Cappellari2026jam} differs fundamentally by accommodating completely general velocity anisotropy. As demonstrated by the modern compilation in \autoref{fig:jam_vs_schw} and \autoref{tab:jam_schw_compilation}, modern JAM and Schwarzschild models applied to comparable data yield BH masses that agree well within their empirical scatter.

\subsection{Validation against mock kinematics and gold-standard masses}
 
In contexts where data are limited or affected by complex spatial systematics, the explicit priors of the JAM framework frequently provide a distinct advantage. Tests using ideal mock IFS kinematics extracted from N-body simulations of a triaxial galaxy \citep[Chapter~5]{Simon2025thesis} reveal that both JAM and Schwarzschild methods recover the true BH mass with comparable accuracy and face similar systematic limitations under highly unfavourable projection angles. Crucially, the simulations show no preference for the non-parametric Schwarzschild approach over the parametric JAM framework. 

These theoretical results corroborate findings from real galaxies where the structural constraints of JAM successfully prevented the overfitting occasionally seen in highly flexible models. For the compact galaxy NGC~1277, an initial Schwarzschild analysis suggested an exceptionally overmassive BH of $M_{\rm BH}=(1.7\pm0.3)\times10^{10}\,\mathrm{M}_\odot$ \citep{vanDenBosch2012}. Re-analysing the same data using the JAM formalism, \citet{Emsellem2013} derived a significantly lower value of $M_{\rm BH}=(5\pm1)\times10^9\,\mathrm{M}_\odot$. This lower mass was later confirmed by both independent Schwarzschild \citep{Walsh2016} and JAM \citep{Krajnovic2018} modelling using higher-resolution adaptive-optics data. Similarly, in NGC~4751, the JAM BH mass \citep{Dominiak2025} lies significantly closer to the independent CO gas-dynamical mass than earlier Schwarzschild models \citep{Rusli2013}.

The absolute accuracy of both methodologies is most definitively demonstrated by their performance on benchmark galaxies where the BH mass is known with exquisite precision from independent methods. In the Milky Way, both JAM and Schwarzschild models applied to the integrated kinematics of the nuclear star cluster successfully recover the true mass derived from individual stellar orbits \citep{FeldmeierKrause2017}. In NGC~4258, which hosts a pristine water-megamaser disc \citep{Miyoshi1995}, both the Schwarzschild analysis of \citet{Siopis2009} and the JAM analysis of \citet{Nguyen2026maser} yield masses within $5-15$ per cent of the geometric value. Likewise, in the giant elliptical M87, stellar-dynamical measurements using both triaxial Schwarzschild models \citep{Gebhardt2011,Liepold2023} and JAM \citep{Simon2024} show excellent agreement with the direct geometric mass measurement obtained from Event Horizon Telescope imaging \citep{EHTC2019}.

\subsection{Compilation Data}

For completeness and reproducibility, \autoref{tab:jam_schw_compilation} lists the $N=33$ published JAM--Schwarzschild measurement pairs used to construct \autoref{fig:jam_vs_schw}. These correspond to 29 distinct stellar systems; multiple entries are retained when independent JAM measurements or materially different JAM configurations are available. The compilation includes measurements obtained from the same dataset as well as comparisons between measurements published in separate studies. When both triaxial and axisymmetric Schwarzschild models are available, we adopt the triaxial result for the statistical comparison. Upper limits are excluded. The preliminary measurements of \citet{Cappellari2010conf} are not included in this table or in the primary statistics.

\begin{table*}
\centering
\caption{Published JAM and Schwarzschild BH mass measurements used in
\autoref{fig:jam_vs_schw}. All masses are in $\mathrm{M}_\odot$.
Different rows for the same object represent independent measurements
or different JAM configurations and are plotted separately.}
\label{tab:jam_schw_compilation}
\begin{tabular}{lcccc}
\hline
Object &
$M_{\rm BH}^{\rm JAM}$ &
JAM reference &
$M_{\rm BH}^{\rm Sch}$ &
Schwarzschild reference \\
\hline
IC~4296 &
$22.0^{+20.2}_{-8.1}\times10^8$ &
\citet{Thater2026} &
$35.1^{+33.7}_{-25.7}\times10^8$ &
\citet{Thater2026} \\

IC~4329 &
$45.8^{+20.3}_{-21.0}\times10^8$ &
\citet{Thater2026} &
$24.3^{+15.3}_{-16.5}\times10^8$ &
\citet{Thater2026} \\

M59-UCD3 &
$5.9^{+2.1}_{-1.7}\times10^6$ &
\citet{Ahn2018} &
$2.5^{+2.1}_{-1.7}\times10^6$ &
\citet{Ahn2018} \\

M60-UCD1 &
$(2.4\pm0.4)\times10^7$ &
\citet{Seth2014} &
$2.1^{+1.4}_{-0.7}\times10^7$ &
\citet{Seth2014} \\

M87 (same photometry) &
$5.5^{+0.5}_{-0.3}\times10^9$ &
\citet{Simon2024} &
$5.3^{+0.37}_{-0.2}\times10^9$ &
\citet{Liepold2023} \\

Milky Way &
$2.8^{+1.3}_{-0.8}\times10^6$ &
\citet{FeldmeierKrause2017} &
$3.0^{+1.1}_{-1.3}\times10^6$ &
\citet{FeldmeierKrause2017} \\

NGC~584 &
$(1.93\pm0.06)\times10^8$ &
\citet{Thater2019} &
$(1.34\pm0.49)\times10^8$ &
\citet{Thater2019} \\

NGC~1277 &
$(4.61\pm0.37)\times10^9$ &
\citet{Krajnovic2018} &
$(4.9\pm0.2)\times10^9$ &
\citet{Walsh2016} \\

NGC~1277 (early data) &
$(5\pm1)\times10^9$ &
\citet{Emsellem2013} &
$(1.7\pm0.3)\times10^{10}$ &
\citet{vanDenBosch2012} \\

NGC~2693 &
$(2.9\pm0.3)\times10^9$ &
\citet{Pilawa2022} &
$(2.4\pm0.6)\times10^9$ &
\citet{Pilawa2022} \\

NGC~2784 &
$(1.69\pm0.10)\times10^8$ &
\citet{Thater2019} &
$(1.03\pm0.54)\times10^8$ &
\citet{Thater2019} \\

NGC~3640 &
$(0.99\pm0.13)\times10^8$ &
\citet{Thater2019} &
$(0.77\pm0.51)\times10^8$ &
\citet{Thater2019} \\

NGC~3706 &
$(5.3\pm0.8)\times10^8$ &
\citet{Thater2026} &
$11.4^{+4.1}_{-6.3}\times10^8$ &
\citet{Thater2026} \\

NGC~3923 &
$10.0^{+4.7}_{-1.8}\times10^8$ &
\citet{Thater2026} &
$11.9^{+13.4}_{-8.0}\times10^8$ &
\citet{Thater2026} \\

NGC~4258 &
$4.8^{+0.8}_{-0.9}\times10^7$ &
\citet{Drehmer2015} &
$(3.3\pm0.2)\times10^7$ &
\citet{Siopis2009} \\

NGC~4258 &
$4.08^{+0.19}_{-0.33}\times10^7$ &
\citet{Nguyen2026maser} &
$(3.3\pm0.2)\times10^7$ &
\citet{Siopis2009} \\

NGC~4261 &
$12.5^{+24.3}_{-7.9}\times10^8$ &
\citet{Thater2026} &
$11.0^{+10.8}_{-9.5}\times10^8$ &
\citet{Thater2026} \\

NGC~4281 &
$(4.91\pm0.15)\times10^8$ &
\citet{Thater2019} &
$(5.42\pm0.80)\times10^8$ &
\citet{Thater2019} \\

NGC~4339 &
$6.5^{+0.7}_{-1.3}\times10^7$ &
\citet{Krajnovic2018} &
$4.3^{+4.8}_{-2.3}\times10^7$ &
\citet{Krajnovic2018} \\

NGC~4434 &
$9.0^{+2.5}_{-0.5}\times10^7$ &
\citet{Krajnovic2018} &
$7.0^{+2.0}_{-2.8}\times10^7$ &
\citet{Krajnovic2018} \\

NGC~4486B &
$3.6\times10^8$ &
\citet{Tahmasebzadeh2025} &
$(3.6\pm0.6)\times10^8$ &
\citet{Tahmasebzadeh2025} \\

NGC~4570 &
$(1.19\pm0.09)\times10^8$ &
\citet{Thater2019} &
$(0.68\pm0.20)\times10^8$ &
\citet{Thater2019} \\

NGC~4578 &
$3.5^{+0.4}_{-0.3}\times10^7$ &
\citet{Krajnovic2018} &
$1.9^{+0.6}_{-1.4}\times10^7$ &
\citet{Krajnovic2018} \\

NGC~4636 &
$7.8^{+4.5}_{-1.7}\times10^8$ &
\citet{Thater2026} &
$4.7^{+3.0}_{-4.3}\times10^8$ &
\citet{Thater2026} \\

NGC~4751 (JAM$_{\rm cyl}$) &
$(2.52\pm0.36)\times10^9$ &
\citet{Dominiak2025} &
$1.4^{+1.5}_{-1.3}\times10^9$ &
\citet{Rusli2013} \\

NGC~4751 (JAM$_{\rm sph}$) &
$(3.24\pm0.87)\times10^9$ &
\citet{Dominiak2025} &
$1.4^{+1.5}_{-1.3}\times10^9$ &
\citet{Rusli2013} \\

NGC~4762 &
$5.2^{+0.9}_{-1.1}\times10^7$ &
\citet{Krajnovic2018} &
$2.3^{+0.9}_{-0.6}\times10^7$ &
\citet{Krajnovic2018} \\

NGC~4889 &
$\simeq2.0\times10^{10}$ &
\citet{Shetty2020} &
$(2.1\pm1.6)\times10^{10}$ &
\citet{McConnell2012} \\

NGC~5102 &
$9.1^{+1.8}_{-1.5}\times10^5$ &
\citet{Nguyen2018, Nguyen2019} &
$1.30^{+0.19}_{-0.18}\times10^6$ &
\citet{Waters2026} \\

NGC~6958 (JAM$_{\rm cyl}$) &
$(8.6\pm0.8)\times10^8$ &
\citet{Thater2022} &
$3.6^{+2.5}_{-1.3}\times10^8$ &
\citet{Thater2022} \\

NGC~6958 (JAM$_{\rm sph}$) &
$4.6^{+2.5}_{-2.7}\times10^8$ &
\citet{Thater2022} &
$3.6^{+2.5}_{-1.3}\times10^8$ &
\citet{Thater2022} \\

NGC~7049 &
$(3.16\pm0.84)\times10^8$ &
\citet{Thater2019} &
$(3.17\pm0.84)\times10^8$ &
\citet{Thater2019} \\

UCD736 &
$1.8^{+1.8}_{-1.4}\times10^6$ &
\citet{Taylor2025} &
$(2.1\pm1.1)\times10^6$ &
\citet{Taylor2025} \\
\hline
\end{tabular}
\end{table*}

The machine-readable version of \autoref{tab:jam_schw_compilation},
including the individual uncertainties, modelling configurations,
upper-limit flags, source references, and the Python script used to
generate \autoref{fig:jam_vs_schw}, will be made available with the
published article.

\label{lastpage}


\begin{thebibliography}{}
\makeatletter
\relax
\def\mn@urlcharsother{\let\do\@makeother \do\$\do\&\do\#\do\^\do\_\do\%\do\~}
\def\mn@doi{\begingroup\mn@urlcharsother \@ifnextchar [ {\mn@doi@}
  {\mn@doi@[]}}
\def\mn@doi@[#1]#2{\def\@tempa{#1}\ifx\@tempa\@empty \href
  {http://dx.doi.org/#2} {doi:#2}\else \href {http://dx.doi.org/#2}
  {\color{blue}#1}\fi \endgroup}
\def\mn@eprint#1#2{\mn@eprint@#1:#2::\@nil}
\def\mn@eprint@arXiv#1{\href {http://arxiv.org/abs/#1} {{\tt arXiv:#1}}}
\def\mn@eprint@dblp#1{\href {http://dblp.uni-trier.de/rec/bibtex/#1.xml}
  {dblp:#1}}
\def\mn@eprint@#1:#2:#3:#4\@nil{\def\@tempa {#1}\def\@tempb {#2}\def\@tempc
  {#3}\ifx \@tempc \@empty \let \@tempc \@tempb \let \@tempb \@tempa \fi \ifx
  \@tempb \@empty \def\@tempb {arXiv}\fi \@ifundefined
  {mn@eprint@\@tempb}{\@tempb:\@tempc}{\expandafter \expandafter \csname
  mn@eprint@\@tempb\endcsname \expandafter{\@tempc}}}

\bibitem[\protect\citeauthoryear{Ahn et~al.,}{Ahn et~al.}{2018}]{Ahn2018}
Ahn C.~P.,  et~al., 2018, \mn@doi [\apj] {10.3847/1538-4357/aabc57}, \href
  {https://ui.adsabs.harvard.edu/abs/2018ApJ...858..102A} {858, 102}

\bibitem[\protect\citeauthoryear{Barth, Sarzi, Rix, Ho, Filippenko  \&
  Sargent}{Barth et~al.}{2001}]{Barth2001}
Barth A.~J.,  Sarzi M.,  Rix H.-W.,  Ho L.~C.,  Filippenko A.~V.,   Sargent W.
  L.~W.,  2001, \mn@doi [\apj] {10.1086/321523}, \href
  {https://ui.adsabs.harvard.edu/abs/2001ApJ...555..685B} {555, 685}

\bibitem[\protect\citeauthoryear{Binney \& Mamon}{Binney \&
  Mamon}{1982}]{Binney1982}
Binney J.,  Mamon G.~A.,  1982, \mn@doi [\mnras] {10.1093/mnras/200.2.361},
  \href {https://ui.adsabs.harvard.edu/abs/1982MNRAS.200..361B} {200, 361}

\bibitem[\protect\citeauthoryear{Binney \& Tremaine}{Binney \&
  Tremaine}{1987}]{Binney1987}
Binney J.,  Tremaine S.,  1987, Galactic dynamics.
Princeton University Press, \url
  {https://books.google.co.uk/books?id=01yNf7mipb0C}

\bibitem[\protect\citeauthoryear{{Blecha}, {Sijacki}  \& {Springel}}{{Blecha}
  et~al.}{2016}]{Blecha2016}
{Blecha} L.,  {Sijacki} D.,   {Springel} V.,  2016, \mn@doi [\mnras]
  {10.1093/mnras/stv2642}, \href
  {https://ui.adsabs.harvard.edu/abs/2016MNRAS.456..961B} {456, 961}

\bibitem[\protect\citeauthoryear{{Campanelli}, {Lousto}, {Zlochower}  \&
  {Merritt}}{{Campanelli} et~al.}{2007}]{Campanelli2007}
{Campanelli} M.,  {Lousto} C.~O.,  {Zlochower} Y.,   {Merritt} D.,  2007,
  \mn@doi [\prl] {10.1103/PhysRevLett.98.231102}, \href
  {https://ui.adsabs.harvard.edu/abs/2007PhRvL..98w1102C} {98, 231102}

\bibitem[\protect\citeauthoryear{Cappellari}{Cappellari}{2002}]{Cappellari2002mge}
Cappellari M.,  2002, \mn@doi [\mnras] {10.1046/j.1365-8711.2002.05412.x},
  \href {https://ui.adsabs.harvard.edu/abs/2002MNRAS.333..400C} {333, 400}

\bibitem[\protect\citeauthoryear{Cappellari}{Cappellari}{2008}]{Cappellari2008}
Cappellari M.,  2008, \mn@doi [\mnras] {10.1111/j.1365-2966.2008.13754.x},
  \href {https://ui.adsabs.harvard.edu/abs/2008MNRAS.390...71C} {390, 71}

\bibitem[\protect\citeauthoryear{Cappellari}{Cappellari}{2016}]{Cappellari2016}
Cappellari M.,  2016, \mn@doi [\araa] {10.1146/annurev-astro-082214-122432},
  \href {https://ui.adsabs.harvard.edu/abs/2016ARA%26A..54..597C} {54, 597}

\bibitem[\protect\citeauthoryear{Cappellari}{Cappellari}{2017}]{Cappellari2017}
Cappellari M.,  2017, \mn@doi [\mnras] {10.1093/mnras/stw3020}, \href
  {https://ui.adsabs.harvard.edu/abs/2017MNRAS.466..798C} {466, 798}

\bibitem[\protect\citeauthoryear{Cappellari}{Cappellari}{2020}]{Cappellari2020}
Cappellari M.,  2020, \mn@doi [\mnras] {10.1093/mnras/staa959}, \href
  {https://ui.adsabs.harvard.edu/abs/2020MNRAS.494.4819C} {494, 4819}

\bibitem[\protect\citeauthoryear{Cappellari}{Cappellari}{2023}]{Cappellari2023}
Cappellari M.,  2023, \mn@doi [\mnras] {10.1093/mnras/stad2597}, \href
  {https://ui.adsabs.harvard.edu/abs/2023MNRAS.526.3273C} {526, 3273}

\bibitem[\protect\citeauthoryear{Cappellari}{Cappellari}{2025}]{Cappellari2025}
Cappellari M.,  2025, \mn@doi [\mnras] {10.1093/mnras/staf1726}, \href
  {https://ui.adsabs.harvard.edu/abs/2025MNRAS.544.1432C} {544, 1432}

\bibitem[\protect\citeauthoryear{Cappellari}{Cappellari}{2026a}]{Cappellari2026}
Cappellari M.,  2026a, in Mandel I.,  ed., , Vol.~4, Encyclopedia of
  Astrophysics, Volume 4.
Elsevier, Amsterdam, The Netherlands, pp 122--152 (\mn@eprint {arXiv}
  {2503.02746}), \mn@doi{10.1016/B978-0-443-21439-4.00109-7}

\bibitem[\protect\citeauthoryear{Cappellari}{Cappellari}{2026b}]{Cappellari2026jam}
Cappellari M.,  2026b, \mn@doi [\mnras] {10.1093/mnras/stag420}, \href
  {https://ui.adsabs.harvard.edu/abs/2026MNRAS.549ag420C} {549, stag420}

\bibitem[\protect\citeauthoryear{Cappellari \& Emsellem}{Cappellari \&
  Emsellem}{2004}]{Cappellari2004}
Cappellari M.,  Emsellem E.,  2004, \mn@doi [\pasp] {10.1086/381875}, \href
  {https://ui.adsabs.harvard.edu/abs/2004PASP..116..138C} {116, 138}

\bibitem[\protect\citeauthoryear{Cappellari et~al.,}{Cappellari
  et~al.}{2010}]{Cappellari2010conf}
Cappellari M.,  et~al., 2010, in {Debattista} V.~P.,  {Popescu} C.~C.,  eds,
  American Institute of Physics Conference Series Vol. 1240, American Institute
  of Physics Conference Series. pp 211--214 (\mn@eprint {arXiv} {1001.3233}),
  \mn@doi{10.1063/1.3458489}

\bibitem[\protect\citeauthoryear{Cappellari et~al.,}{Cappellari
  et~al.}{2013a}]{Cappellari2013p15}
Cappellari M.,  et~al., 2013a, \mn@doi [\mnras] {10.1093/mnras/stt562}, \href
  {https://ui.adsabs.harvard.edu/abs/2013MNRAS.432.1709C} {432, 1709}

\bibitem[\protect\citeauthoryear{Cappellari et~al.,}{Cappellari
  et~al.}{2013b}]{Cappellari2013p20}
Cappellari M.,  et~al., 2013b, \mn@doi [\mnras] {10.1093/mnras/stt644}, \href
  {https://ui.adsabs.harvard.edu/abs/2013MNRAS.432.1862C} {432, 1862}

\bibitem[\protect\citeauthoryear{{Cappellari}, {Nguyen}  \&
  {Kassin}}{{Cappellari} et~al.}{2026b}]{Cappellari2026p3}
{Cappellari} M.,  {Nguyen} D.~D.,   {Kassin} S.~A.,  2026b, arXiv, submitted

\bibitem[\protect\citeauthoryear{{Cappellari}, {Nguyen}  \&
  {Kassin}}{{Cappellari} et~al.}{2026a}]{Cappellari2026p1}
{Cappellari} M.,  {Nguyen} D.~D.,   {Kassin} S.~A.,  2026a, \mnras, submitted

\bibitem[\protect\citeauthoryear{Davis, Bureau, Cappellari, Sarzi  \&
  Blitz}{Davis et~al.}{2013}]{Davis2013}
Davis T.~A.,  Bureau M.,  Cappellari M.,  Sarzi M.,   Blitz L.,  2013, \mn@doi
  [\nat] {10.1038/nature11819}, \href
  {https://ui.adsabs.harvard.edu/abs/2013Natur.494..328D} {494, 328}

\bibitem[\protect\citeauthoryear{Dey et~al.,}{Dey et~al.}{2019}]{Dey2019}
Dey A.,  et~al., 2019, \mn@doi [\aj] {10.3847/1538-3881/ab089d}, \href
  {https://ui.adsabs.harvard.edu/abs/2019AJ....157..168D} {157, 168}

\bibitem[\protect\citeauthoryear{{Dominiak} et~al.,}{{Dominiak}
  et~al.}{2025}]{Dominiak2025}
{Dominiak} P.,  et~al., 2025, \mn@doi [\mnras] {10.1093/mnras/staf1338}, \href
  {https://ui.adsabs.harvard.edu/abs/2025MNRAS.542.2039D} {542, 2039}

\bibitem[\protect\citeauthoryear{Drehmer, Storchi-Bergmann, Ferrari, Cappellari
   \& Riffel}{Drehmer et~al.}{2015}]{Drehmer2015}
Drehmer D.~A.,  Storchi-Bergmann T.,  Ferrari F.,  Cappellari M.,   Riffel
  R.~A.,  2015, \mn@doi [\mnras] {10.1093/mnras/stv536}, \href
  {https://ui.adsabs.harvard.edu/abs/2015MNRAS.450..128D} {450, 128}

\bibitem[\protect\citeauthoryear{Emsellem}{Emsellem}{2013}]{Emsellem2013}
Emsellem E.,  2013, \mn@doi [\mnras] {10.1093/mnras/stt840}, \href
  {https://ui.adsabs.harvard.edu/abs/2013MNRAS.433.1862E} {433, 1862}

\bibitem[\protect\citeauthoryear{Emsellem, Monnet  \& Bacon}{Emsellem
  et~al.}{1994}]{Emsellem1994}
Emsellem E.,  Monnet G.,   Bacon R.,  1994, \aap, \href
  {https://ui.adsabs.harvard.edu/abs/1994A%26A...285..723E} {285, 723}

\bibitem[\protect\citeauthoryear{Emsellem et~al.,}{Emsellem
  et~al.}{2011}]{Emsellem2011p3}
Emsellem E.,  et~al., 2011, \mn@doi [\mnras]
  {10.1111/j.1365-2966.2011.18496.x}, \href
  {https://ui.adsabs.harvard.edu/abs/2011MNRAS.414..888E} {414, 888}

\bibitem[\protect\citeauthoryear{{Event Horizon Telescope Collaboration}
  et~al.,}{{Event Horizon Telescope Collaboration} et~al.}{2019}]{EHTC2019}
{Event Horizon Telescope Collaboration} et~al., 2019, \mn@doi [\apjl]
  {10.3847/2041-8213/ab1141}, \href
  {https://ui.adsabs.harvard.edu/abs/2019ApJ...875L...6E} {875, L6}

\bibitem[\protect\citeauthoryear{Feldmeier-Krause, Zhu, Neumayer, van~de Ven,
  de Zeeuw  \& Sch{\"o}del}{Feldmeier-Krause
  et~al.}{2017}]{FeldmeierKrause2017}
Feldmeier-Krause A.,  Zhu L.,  Neumayer N.,  van~de Ven G.,  de Zeeuw P.~T.,
  Sch{\"o}del R.,  2017, \mn@doi [\mnras] {10.1093/mnras/stw3377}, \href
  {https://ui.adsabs.harvard.edu/abs/2017MNRAS.466.4040F} {466, 4040}

\bibitem[\protect\citeauthoryear{Ferrarese \& Merritt}{Ferrarese \&
  Merritt}{2000}]{Ferrarese2000}
Ferrarese L.,  Merritt D.,  2000, \mn@doi [\apjl] {10.1086/312838}, \href
  {https://ui.adsabs.harvard.edu/abs/2000ApJ...539L...9F} {539, L9}

\bibitem[\protect\citeauthoryear{{Gebhardt}}{{Gebhardt}}{2004}]{Gebhardt2004}
{Gebhardt} K.,  2004, in {Ho} L.~C.,  ed., Coevolution of Black Holes and
  Galaxies. p.~248 (\mn@eprint {arXiv} {astro-ph/0306090}),
  \mn@doi{10.48550/arXiv.astro-ph/0306090}

\bibitem[\protect\citeauthoryear{Gebhardt et~al.,}{Gebhardt
  et~al.}{2000}]{Gebhardt2000bh}
Gebhardt K.,  et~al., 2000, \mn@doi [\apjl] {10.1086/312840}, \href
  {https://ui.adsabs.harvard.edu/abs/2000ApJ...539L..13G} {539, L13}

\bibitem[\protect\citeauthoryear{Gebhardt, Adams, Richstone, Lauer, Faber,
  G{\"u}ltekin, Murphy  \& Tremaine}{Gebhardt et~al.}{2011}]{Gebhardt2011}
Gebhardt K.,  Adams J.,  Richstone D.,  Lauer T.~R.,  Faber S.~M.,
  G{\"u}ltekin K.,  Murphy J.,   Tremaine S.,  2011, \mn@doi [\apj]
  {10.1088/0004-637X/729/2/119}, \href
  {https://ui.adsabs.harvard.edu/abs/2011ApJ...729..119G} {729, 119}

\bibitem[\protect\citeauthoryear{Gerhard}{Gerhard}{1993}]{Gerhard1993}
Gerhard O.~E.,  1993, \mn@doi [\mnras] {10.1093/mnras/265.1.213}, \href
  {https://ui.adsabs.harvard.edu/abs/1993MNRAS.265..213G} {265, 213}

\bibitem[\protect\citeauthoryear{Gillis}{Gillis}{2014}]{Gillis2014}
Gillis N.,  2014, \mn@doi [SIAM Journal on Imaging Sciences]
  {10.1137/130946782}, 7, 1420

\bibitem[\protect\citeauthoryear{Gillis}{Gillis}{2020}]{Gillis2020}
Gillis N.,  2020, {Nonnegative Matrix Factorization}.
Society for Industrial and Applied Mathematics (SIAM), Philadelphia, PA,
  \mn@doi{10.1137/1.9781611976410}

\bibitem[\protect\citeauthoryear{{Gualandris} \& {Merritt}}{{Gualandris} \&
  {Merritt}}{2008}]{Gualandris2008}
{Gualandris} A.,  {Merritt} D.,  2008, \mn@doi [\apj] {10.1086/586877}, \href
  {https://ui.adsabs.harvard.edu/abs/2008ApJ...678..780G} {678, 780}

\bibitem[\protect\citeauthoryear{Haario, Saksman  \& Tamminen}{Haario
  et~al.}{2001}]{Haario2001}
Haario H.,  Saksman E.,   Tamminen J.,  2001, \mn@doi [Bernoulli]
  {10.2307/3318737}, 7, 223

\bibitem[\protect\citeauthoryear{H{\"a}ring \& Rix}{H{\"a}ring \&
  Rix}{2004}]{Haring2004}
H{\"a}ring N.,  Rix H.-W.,  2004, \mn@doi [\apjl] {10.1086/383567}, \href
  {https://ui.adsabs.harvard.edu/abs/2004ApJ...604L..89H} {604, L89}

\bibitem[\protect\citeauthoryear{Harms et~al.,}{Harms et~al.}{1994}]{Harms1994}
Harms R.~J.,  et~al., 1994, \mn@doi [\apjl] {10.1086/187588}, \href
  {https://ui.adsabs.harvard.edu/abs/1994ApJ...435L..35H} {435, L35}

\bibitem[\protect\citeauthoryear{{Hoffman} \& {Loeb}}{{Hoffman} \&
  {Loeb}}{2007}]{Hoffman2007}
{Hoffman} L.,  {Loeb} A.,  2007, \mn@doi [\mnras]
  {10.1111/j.1365-2966.2007.11694.x}, \href
  {https://ui.adsabs.harvard.edu/abs/2007MNRAS.377..957H} {377, 957}

\bibitem[\protect\citeauthoryear{Jeans}{Jeans}{1915}]{Jeans1915}
Jeans J.~H.,  1915, \mn@doi [\mnras] {10.1093/mnras/76.2.70}, \href
  {https://ui.adsabs.harvard.edu/abs/1915MNRAS..76...70J} {76, 70}

\bibitem[\protect\citeauthoryear{Jin, Zhu, Long, Mao, Xu, Li  \& van~de
  Ven}{Jin et~al.}{2019}]{Jin2019}
Jin Y.,  Zhu L.,  Long R.~J.,  Mao S.,  Xu D.,  Li H.,   van~de Ven G.,  2019,
  \mn@doi [\mnras] {10.1093/mnras/stz1170}, \href
  {https://ui.adsabs.harvard.edu/abs/2019MNRAS.486.4753J} {486, 4753}

\bibitem[\protect\citeauthoryear{{Komossa}}{{Komossa}}{2012}]{Komossa2012}
{Komossa} S.,  2012, \mn@doi [Advances in Astronomy] {10.1155/2012/364927},
  \href {https://ui.adsabs.harvard.edu/abs/2012AdAst2012E..14K} {2012, 364927}

\bibitem[\protect\citeauthoryear{Kormendy \& Ho}{Kormendy \&
  Ho}{2013}]{Kormendy2013review}
Kormendy J.,  Ho L.~C.,  2013, \mn@doi [\araa]
  {10.1146/annurev-astro-082708-101811}, \href
  {https://ui.adsabs.harvard.edu/abs/2013ARA%26A..51..511K} {51, 511}

\bibitem[\protect\citeauthoryear{Krajnovi{\'c} et~al.,}{Krajnovi{\'c}
  et~al.}{2018}]{Krajnovic2018}
Krajnovi{\'c} D.,  et~al., 2018, \mn@doi [\mnras] {10.1093/mnras/sty778}, \href
  {https://ui.adsabs.harvard.edu/abs/2018MNRAS.477.3030K} {477, 3030}

\bibitem[\protect\citeauthoryear{{Krajnovi{\'c}}, {Emsellem}, {Bacon},
  {Boogaard}, {Weilbacher}  \& {Wisotzki}}{{Krajnovi{\'c}}
  et~al.}{2026}]{Krajnovic2026}
{Krajnovi{\'c}} D.,  {Emsellem} E.,  {Bacon} R.,  {Boogaard} L.~A.,
  {Weilbacher} P.~M.,   {Wisotzki} L.,  2026, \mn@doi [\aap]
  {10.1051/0004-6361/202558624}, \href
  {https://ui.adsabs.harvard.edu/abs/2026A&A...708A..15K} {708, A15}

\bibitem[\protect\citeauthoryear{Krajnovi{\'c et al.}}{Krajnovi{\'c et
  al.}}{2018a}]{Krajnovic2018channels}
Krajnovi{\'c et al.} 2018a, \mn@doi [\mnras] {10.1093/mnras/stx2704}, \href
  {https://ui.adsabs.harvard.edu/abs/2018MNRAS.473.5237K} {473, 5237}

\bibitem[\protect\citeauthoryear{Krajnovi{\'c et al.}}{Krajnovi{\'c et
  al.}}{2018b}]{Krajnovic2018prolate}
Krajnovi{\'c et al.} 2018b, \mn@doi [\mnras] {10.1093/mnras/sty1031}, \href
  {https://ui.adsabs.harvard.edu/abs/2018MNRAS.477.5327K} {477, 5327}

\bibitem[\protect\citeauthoryear{Leung et~al.,}{Leung et~al.}{2018}]{Leung2018}
Leung G. Y.~C.,  et~al., 2018, \mn@doi [\mnras] {10.1093/mnras/sty288}, \href
  {https://ui.adsabs.harvard.edu/abs/2018MNRAS.477..254L} {477, 254}

\bibitem[\protect\citeauthoryear{Li, Mao, Cappellari, Graham, Emsellem  \&
  Long}{Li et~al.}{2018}]{Li2018shapes}
Li H.,  Mao S.,  Cappellari M.,  Graham M.~T.,  Emsellem E.,   Long R.~J.,
  2018, \mn@doi [\apjl] {10.3847/2041-8213/aad54b}, \href
  {https://ui.adsabs.harvard.edu/abs/2018ApJ...863L..19L} {863, L19}

\bibitem[\protect\citeauthoryear{Liepold, Ma  \& Walsh}{Liepold
  et~al.}{2023}]{Liepold2023}
Liepold E.~R.,  Ma C.-P.,   Walsh J.~L.,  2023, \mn@doi [\mnras]
  {10.1093/mnras/stad2423}, 525, 4971

\bibitem[\protect\citeauthoryear{Macchetto, Marconi, Axon, Capetti, Sparks  \&
  Crane}{Macchetto et~al.}{1997}]{Macchetto1997}
Macchetto F.,  Marconi A.,  Axon D.~J.,  Capetti A.,  Sparks W.,   Crane P.,
  1997, \mn@doi [\apj] {10.1086/304823}, \href
  {https://ui.adsabs.harvard.edu/abs/1997ApJ...489..579M} {489, 579}

\bibitem[\protect\citeauthoryear{Magorrian et~al.,}{Magorrian
  et~al.}{1998}]{Magorrian1998}
Magorrian J.,  et~al., 1998, \mn@doi [\aj] {10.1086/300353}, \href
  {https://ui.adsabs.harvard.edu/abs/1998AJ....115.2285M} {115, 2285}

\bibitem[\protect\citeauthoryear{Marconi \& Hunt}{Marconi \&
  Hunt}{2003}]{Marconi2003}
Marconi A.,  Hunt L.~K.,  2003, \mn@doi [\apjl] {10.1086/375804}, \href
  {https://ui.adsabs.harvard.edu/abs/2003ApJ...589L..21M} {589, L21}

\bibitem[\protect\citeauthoryear{McConnell, Ma, Murphy, Gebhardt, Lauer,
  Graham, Wright  \& Richstone}{McConnell et~al.}{2012}]{McConnell2012}
McConnell N.~J.,  Ma C.-P.,  Murphy J.~D.,  Gebhardt K.,  Lauer T.~R.,  Graham
  J.~R.,  Wright S.~A.,   Richstone D.~O.,  2012, \mn@doi [\apj]
  {10.1088/0004-637X/756/2/179}, \href
  {https://ui.adsabs.harvard.edu/abs/2012ApJ...756..179M} {756, 179}

\bibitem[\protect\citeauthoryear{Merritt \& Milosavljevi{\'c}}{Merritt \&
  Milosavljevi{\'c}}{2005}]{Merritt2005}
Merritt D.,  Milosavljevi{\'c} M.,  2005, Living Reviews in Relativity, \href
  {https://ui.adsabs.harvard.edu/abs/2005LRR.....8....8M} {8, 8}

\bibitem[\protect\citeauthoryear{M{\'e}sz{\'a}ros et~al.,}{M{\'e}sz{\'a}ros
  et~al.}{2024}]{Meszaros2024}
M{\'e}sz{\'a}ros S.,  et~al., 2024, \mn@doi [\aap]
  {10.1051/0004-6361/202449306}, \href
  {https://ui.adsabs.harvard.edu/abs/2024A&A...688A.197M} {688, A197}

\bibitem[\protect\citeauthoryear{Mitzkus, Cappellari  \& Walcher}{Mitzkus
  et~al.}{2017}]{Mitzkus2017}
Mitzkus M.,  Cappellari M.,   Walcher C.~J.,  2017, \mn@doi [\mnras]
  {10.1093/mnras/stw2677}, \href
  {https://ui.adsabs.harvard.edu/abs/2017MNRAS.464.4789M} {464, 4789}

\bibitem[\protect\citeauthoryear{Miyoshi, Moran, Herrnstein, Greenhill, Nakai,
  Diamond  \& Inoue}{Miyoshi et~al.}{1995}]{Miyoshi1995}
Miyoshi M.,  Moran J.,  Herrnstein J.,  Greenhill L.,  Nakai N.,  Diamond P.,
  Inoue M.,  1995, \mn@doi [\nat] {10.1038/373127a0}, \href
  {https://ui.adsabs.harvard.edu/abs/1995Natur.373..127M} {373, 127}

\bibitem[\protect\citeauthoryear{Naab \& Ostriker}{Naab \&
  Ostriker}{2017}]{Naab2017}
Naab T.,  Ostriker J.~P.,  2017, \mn@doi [\araa]
  {10.1146/annurev-astro-081913-040019}, \href
  {https://ui.adsabs.harvard.edu/abs/2017ARA&A..55...59N} {55, 59}

\bibitem[\protect\citeauthoryear{{Neureiter} et~al.,}{{Neureiter}
  et~al.}{2021}]{Neureiter2021}
{Neureiter} B.,  et~al., 2021, \mn@doi [\mnras] {10.1093/mnras/staa3014}, \href
  {https://ui.adsabs.harvard.edu/abs/2021MNRAS.500.1437N} {500, 1437}

\bibitem[\protect\citeauthoryear{{Ngo} et~al.,}{{Ngo} et~al.}{2025}]{Ngo2025}
{Ngo} H.~N.,  et~al., 2025, \mn@doi [\apj] {10.3847/1538-4357/ae0455}, \href
  {https://ui.adsabs.harvard.edu/abs/2025ApJ...992..211N} {992, 211}

\bibitem[\protect\citeauthoryear{Nguyen et~al.,}{Nguyen
  et~al.}{2018}]{Nguyen2018}
Nguyen D.~D.,  et~al., 2018, \mn@doi [\apj] {10.3847/1538-4357/aabe28}, \href
  {https://ui.adsabs.harvard.edu/abs/2018ApJ...858..118N} {858, 118}

\bibitem[\protect\citeauthoryear{Nguyen et~al.,}{Nguyen
  et~al.}{2019}]{Nguyen2019}
Nguyen D.~D.,  et~al., 2019, \mn@doi [\apj] {10.3847/1538-4357/aafe7a}, \href
  {https://ui.adsabs.harvard.edu/abs/2019ApJ...872..104N} {872, 104}

\bibitem[\protect\citeauthoryear{{Nguyen} et~al.,}{{Nguyen}
  et~al.}{2020}]{Nguyen2020}
{Nguyen} D.~D.,  et~al., 2020, \mn@doi [\apj] {10.3847/1538-4357/ab77aa}, \href
  {https://ui.adsabs.harvard.edu/abs/2020ApJ...892...68N} {892, 68}

\bibitem[\protect\citeauthoryear{{Nguyen} et~al.,}{{Nguyen}
  et~al.}{2021}]{Nguyen2021}
{Nguyen} D.~D.,  et~al., 2021, \mn@doi [\mnras] {10.1093/mnras/stab1002}, \href
  {https://ui.adsabs.harvard.edu/abs/2021MNRAS.504.4123N} {504, 4123}

\bibitem[\protect\citeauthoryear{{Nguyen} et~al.,}{{Nguyen}
  et~al.}{2022}]{Nguyen2022}
{Nguyen} D.~D.,  et~al., 2022, \mn@doi [\mnras] {10.1093/mnras/stab3016}, \href
  {https://ui.adsabs.harvard.edu/abs/2022MNRAS.509.2920N} {509, 2920}

\bibitem[\protect\citeauthoryear{Nguyen, Cappellari  \&
  Pereira-Santaella}{Nguyen et~al.}{2023}]{Nguyen2023}
Nguyen D.~D.,  Cappellari M.,   Pereira-Santaella M.,  2023, \mn@doi [\mnras]
  {10.1093/mnras/stad2860}, \href
  {https://ui.adsabs.harvard.edu/abs/2023MNRAS.526.3548N} {526, 3548}

\bibitem[\protect\citeauthoryear{{Nguyen} et~al.,}{{Nguyen}
  et~al.}{2026a}]{Nguyen2026_315}
{Nguyen} D.~D.,  et~al., 2026a, \mn@doi [arXiv e-prints]
  {10.48550/arXiv.2608.31015}, \href
  {https://ui.adsabs.harvard.edu/abs/2026arXiv260831015N} {p. arXiv:2608.31015}

\bibitem[\protect\citeauthoryear{Nguyen et~al.,}{Nguyen
  et~al.}{2026b}]{Nguyen2026maser}
Nguyen D.~D.,  et~al., 2026b, \mn@doi [The Astrophysical Journal]
  {10.3847/1538-4357/ae32f2}, \href
  {https://ui.adsabs.harvard.edu/abs/2025arXiv250920519N} {999, 97}

\bibitem[\protect\citeauthoryear{Nguyen et~al.,}{Nguyen
  et~al.}{2026c}]{Nguyen2026m81}
Nguyen D.~D.,  et~al., 2026c, \mn@doi [\apj] {10.3847/1538-4357/ae64f0}, \href
  {https://ui.adsabs.harvard.edu/abs/2026ApJ..1003...98N} {1003, 98}

\bibitem[\protect\citeauthoryear{{Nguyen} et~al.,}{{Nguyen}
  et~al.}{2026d}]{Nguyen2026_4061}
{Nguyen} D.~D.,  et~al., 2026d, \mn@doi [\apj] {10.3847/1538-4357/ae771c},
  \href {https://ui.adsabs.harvard.edu/abs/2026ApJ..1005...94N} {1005, 94}

\bibitem[\protect\citeauthoryear{Oser, Ostriker, Naab, Johansson  \&
  Burkert}{Oser et~al.}{2010}]{Oser2010}
Oser L.,  Ostriker J.~P.,  Naab T.,  Johansson P.~H.,   Burkert A.,  2010,
  \mn@doi [\apj] {10.1088/0004-637X/725/2/2312}, \href
  {https://ui.adsabs.harvard.edu/abs/2010ApJ...725.2312O} {725, 2312}

\bibitem[\protect\citeauthoryear{Pilawa, Liepold, Ma, Walsh, Quenneville,
  Greene, Thomas  \& Blakeslee}{Pilawa et~al.}{2022}]{Pilawa2022}
Pilawa J.~D.,  Liepold E.~R.,  Ma C.-P.,  Walsh J.~L.,  Quenneville M.~E.,
  Greene J.~E.,  Thomas J.,   Blakeslee J.~P.,  2022, \mn@doi [\apj]
  {10.3847/1538-4357/ac58e2}, 931, 133

\bibitem[\protect\citeauthoryear{{Quenneville}, {Liepold}  \&
  {Ma}}{{Quenneville} et~al.}{2021}]{Quenneville2021}
{Quenneville} M.~E.,  {Liepold} E.~R.,   {Ma} C.-P.,  2021, \mn@doi [\apjs]
  {10.3847/1538-4365/abe6a0}, \href
  {https://ui.adsabs.harvard.edu/abs/2021ApJS..254...25Q} {254, 25}

\bibitem[\protect\citeauthoryear{Rusli et~al.,}{Rusli et~al.}{2013}]{Rusli2013}
Rusli S.~P.,  et~al., 2013, \mn@doi [\aj] {10.1088/0004-6256/146/3/45}, \href
  {https://ui.adsabs.harvard.edu/abs/2013AJ....146...45R} {146, 45}

\bibitem[\protect\citeauthoryear{Schwarzschild}{Schwarzschild}{1979}]{Schwarzschild1979}
Schwarzschild M.,  1979, \mn@doi [\apj] {10.1086/157282}, \href
  {https://ui.adsabs.harvard.edu/abs/1979ApJ...232..236S} {232, 236}

\bibitem[\protect\citeauthoryear{Seth et~al.,}{Seth et~al.}{2014}]{Seth2014}
Seth A.~C.,  et~al., 2014, \mn@doi [\nat] {10.1038/nature13762}, \href
  {https://ui.adsabs.harvard.edu/abs/2014Natur.513..398S} {513, 398}

\bibitem[\protect\citeauthoryear{Shetty, Cappellari, McDermid, Krajnovi{\'c},
  de Zeeuw, Davies  \& Kobayashi}{Shetty et~al.}{2020}]{Shetty2020}
Shetty S.,  Cappellari M.,  McDermid R.~M.,  Krajnovi{\'c} D.,  de Zeeuw P.~T.,
   Davies R.~L.,   Kobayashi C.,  2020, \mn@doi [\mnras]
  {10.1093/mnras/staa1043}, \href
  {https://ui.adsabs.harvard.edu/abs/2020MNRAS.494.5619S} {494, 5619}

\bibitem[\protect\citeauthoryear{{Simon}}{{Simon}}{2025}]{Simon2025thesis}
{Simon} D.~A.,  2025, PhD thesis, University of Oxford, \url
  {https://ora.ox.ac.uk/objects/uuid:25f8b6f9-57fe-41fe-b259-75b22bca4e91}

\bibitem[\protect\citeauthoryear{Simon, Cappellari  \& Hartke}{Simon
  et~al.}{2024}]{Simon2024}
Simon D.~A.,  Cappellari M.,   Hartke J.,  2024, \mn@doi [\mnras]
  {10.1093/mnras/stad3309}, \href
  {https://ui.adsabs.harvard.edu/abs/2024MNRAS.527.2341S} {527, 2341}

\bibitem[\protect\citeauthoryear{Siopis et~al.,}{Siopis
  et~al.}{2009}]{Siopis2009}
Siopis C.,  et~al., 2009, \mn@doi [\apj] {10.1088/0004-637X/693/1/946}, \href
  {https://ui.adsabs.harvard.edu/abs/2009ApJ...693..946S} {693, 946}

\bibitem[\protect\citeauthoryear{{Smith} et~al.,}{{Smith}
  et~al.}{2019}]{Smith2019}
{Smith} M.~D.,  et~al., 2019, \mn@doi [\mnras] {10.1093/mnras/stz625}, \href
  {https://ui.adsabs.harvard.edu/abs/2019MNRAS.485.4359S} {485, 4359}

\bibitem[\protect\citeauthoryear{{Smith} et~al.,}{{Smith}
  et~al.}{2021}]{Smith2021}
{Smith} M.~D.,  et~al., 2021, \mn@doi [\mnras] {10.1093/mnras/stab791}, \href
  {https://ui.adsabs.harvard.edu/abs/2021MNRAS.503.5984S} {503, 5984}

\bibitem[\protect\citeauthoryear{Tahmasebzadeh et~al.,}{Tahmasebzadeh
  et~al.}{2025}]{Tahmasebzadeh2025}
Tahmasebzadeh B.,  et~al., 2025, \mn@doi [\apjl] {10.3847/2041-8213/adf728},
  \href {https://ui.adsabs.harvard.edu/abs/2025ApJ...989L..42T} {989, L42}

\bibitem[\protect\citeauthoryear{Taylor et~al.,}{Taylor
  et~al.}{2025}]{Taylor2025}
Taylor M.~A.,  et~al., 2025, \mn@doi [\apjl] {10.3847/2041-8213/ae028e}, \href
  {https://ui.adsabs.harvard.edu/abs/2025ApJ...991L..24T} {991, L24}

\bibitem[\protect\citeauthoryear{{Thater}, Krajnovi{\'c}, Cappellari, Davis, de
  Zeeuw, McDermid  \& Sarzi}{{Thater} et~al.}{2019}]{Thater2019}
{Thater} S.,  Krajnovi{\'c} D.,  Cappellari M.,  Davis T.~A.,  de Zeeuw P.~T.,
  McDermid R.~M.,   Sarzi M.,  2019, \mn@doi [\aap]
  {10.1051/0004-6361/201834808}, \href
  {https://ui.adsabs.harvard.edu/abs/2019A&A...625A..62T} {625, A62}

\bibitem[\protect\citeauthoryear{{Thater} et~al.,}{{Thater}
  et~al.}{2022a}]{Thater2022}
{Thater} S.,  et~al., 2022a, \mn@doi [\mnras] {10.1093/mnras/stab3210}, \href
  {https://ui.adsabs.harvard.edu/abs/2022MNRAS.509.5416T} {509, 5416}

\bibitem[\protect\citeauthoryear{{Thater} et~al.,}{{Thater}
  et~al.}{2022b}]{Thater2022dynamite}
{Thater} S.,  et~al., 2022b, \mn@doi [\aap] {10.1051/0004-6361/202243926},
  \href {https://ui.adsabs.harvard.edu/abs/2022A&A...667A..51T} {667, A51}

\bibitem[\protect\citeauthoryear{{Thater}, {Chaturvedi}, {Krajnovi{\'c}},
  {Cappellari}, {Khochfar}, {Naab}, {Sarzi}  \& {van de Ven}}{{Thater}
  et~al.}{2026}]{Thater2026}
{Thater} S.,  {Chaturvedi} A.,  {Krajnovi{\'c}} D.,  {Cappellari} M.,
  {Khochfar} S.,  {Naab} T.,  {Sarzi} M.,   {van de Ven} G.,  2026, \mn@doi
  [\aap] {10.1051/0004-6361/202557888}, \href
  {https://ui.adsabs.harvard.edu/abs/2026A&A...712A.105T} {712, A105}

\bibitem[\protect\citeauthoryear{Tremaine, Richstone, Byun, Dressler, Faber,
  Grillmair, Kormendy  \& Lauer}{Tremaine et~al.}{1994}]{Tremaine1994}
Tremaine S.,  Richstone D.~O.,  Byun Y.-I.,  Dressler A.,  Faber S.~M.,
  Grillmair C.,  Kormendy J.,   Lauer T.~R.,  1994, \mn@doi [\aj]
  {10.1086/116883}, \href
  {https://ui.adsabs.harvard.edu/abs/1994AJ....107..634T} {107, 634}

\bibitem[\protect\citeauthoryear{{Vasiliev} \& {Valluri}}{{Vasiliev} \&
  {Valluri}}{2020}]{Vasiliev2020}
{Vasiliev} E.,  {Valluri} M.,  2020, \mn@doi [\apj] {10.3847/1538-4357/ab5fe0},
  \href {https://ui.adsabs.harvard.edu/abs/2020ApJ...889...39V} {889, 39}

\bibitem[\protect\citeauthoryear{Verolme et~al.,}{Verolme
  et~al.}{2002}]{Verolme2002}
Verolme E.~K.,  et~al., 2002, \mn@doi [\mnras]
  {10.1046/j.1365-8711.2002.05664.x}, \href
  {https://ui.adsabs.harvard.edu/abs/2002MNRAS.335..517V} {335, 517}

\bibitem[\protect\citeauthoryear{Walsh, van~den Bosch, Gebhardt,
  Y{\i}ld{\i}r{\i}m, Richstone, G{\"u}ltekin  \& Husemann}{Walsh
  et~al.}{2016}]{Walsh2016}
Walsh J.~L.,  van~den Bosch R. C.~E.,  Gebhardt K.,  Y{\i}ld{\i}r{\i}m A.,
  Richstone D.~O.,  G{\"u}ltekin K.,   Husemann B.,  2016, \mn@doi [\apj]
  {10.3847/0004-637X/817/1/2}, \href
  {https://ui.adsabs.harvard.edu/abs/2016ApJ...817....2W} {817, 2}

\bibitem[\protect\citeauthoryear{Waters, Seth, Nguyen, Neumayer, Cappellari  \&
  Walsh}{Waters et~al.}{2026}]{Waters2026}
Waters C.,  Seth A.~C.,  Nguyen D.~D.,  Neumayer N.,  Cappellari M.,   Walsh
  J.~L.,  2026, \apj, in press

\bibitem[\protect\citeauthoryear{Weijmans et~al.,}{Weijmans
  et~al.}{2014}]{Weijmans2014p24}
Weijmans A.-M.,  et~al., 2014, \mn@doi [\mnras] {10.1093/mnras/stu1603}, \href
  {https://ui.adsabs.harvard.edu/abs/2014MNRAS.444.3340W} {444, 3340}

\bibitem[\protect\citeauthoryear{{Zhang} et~al.,}{{Zhang}
  et~al.}{2025}]{Zhang2025}
{Zhang} H.,  et~al., 2025, \mn@doi [\mnras] {10.1093/mnras/staf055}, \href
  {https://ui.adsabs.harvard.edu/abs/2025MNRAS.537..520Z} {537, 520}

\bibitem[\protect\citeauthoryear{{Zhu} et~al.,}{{Zhu}
  et~al.}{2020}]{LingZhu2020}
{Zhu} L.,  et~al., 2020, \mn@doi [\mnras] {10.1093/mnras/staa1584}, \href
  {https://ui.adsabs.harvard.edu/abs/2020MNRAS.496.1579Z} {496, 1579}

\bibitem[\protect\citeauthoryear{{van Dokkum} et~al.,}{{van Dokkum}
  et~al.}{2026}]{vanDokkum2026}
{van Dokkum} P.,  et~al., 2026, \mn@doi [\apjl] {10.3847/2041-8213/ae3d0e},
  \href {https://ui.adsabs.harvard.edu/abs/2026ApJ...998L..27V} {998, L27}

\bibitem[\protect\citeauthoryear{van~den Bosch \& van~de Ven}{van~den Bosch \&
  van~de Ven}{2009}]{vandenBosch2009}
van~den Bosch R. C.~E.,  van~de Ven G.,  2009, \mn@doi [\mnras]
  {10.1111/j.1365-2966.2009.15177.x}, \href
  {https://ui.adsabs.harvard.edu/abs/2009MNRAS.398.1117V} {398, 1117}

\bibitem[\protect\citeauthoryear{van~den Bosch, van~de Ven, Verolme, Cappellari
   \& de Zeeuw}{van~den Bosch et~al.}{2008}]{vandenBosch2008}
van~den Bosch R. C.~E.,  van~de Ven G.,  Verolme E.~K.,  Cappellari M.,   de
  Zeeuw P.~T.,  2008, \mn@doi [\mnras] {10.1111/j.1365-2966.2008.12874.x},
  \href {https://ui.adsabs.harvard.edu/abs/2008MNRAS.385..647V} {385, 647}

\bibitem[\protect\citeauthoryear{van~den Bosch, Gebhardt, G{\"u}ltekin, van~de
  Ven, van~der Wel  \& Walsh}{van~den Bosch et~al.}{2012}]{vanDenBosch2012}
van~den Bosch R. C.~E.,  Gebhardt K.,  G{\"u}ltekin K.,  van~de Ven G.,
  van~der Wel A.,   Walsh J.~L.,  2012, \mn@doi [\nat] {10.1038/nature11592},
  \href {https://ui.adsabs.harvard.edu/abs/2012Natur.491..729V} {491, 729}

\bibitem[\protect\citeauthoryear{van~der Marel, Cretton, de Zeeuw  \&
  Rix}{van~der Marel et~al.}{1998}]{vanderMarel1998}
van~der Marel R.~P.,  Cretton N.,  de Zeeuw P.~T.,   Rix H.-W.,  1998, \mn@doi
  [\apj] {10.1086/305147}, \href
  {https://ui.adsabs.harvard.edu/abs/1998ApJ...493..613V} {493, 613}

\makeatother
\end{thebibliography}
\end{document}